\documentclass[a4paper,journal]{IEEEtran}
\usepackage{graphicx,ifthen}
\usepackage[cmex10]{amsmath}
\usepackage{psfrag,amssymb}
\usepackage{mathtools}

\usepackage[noadjust]{cite}

\newtheorem{thm}{Theorem}
\newtheorem{lem}[thm]{Lemma}
\newtheorem{cor}[thm]{Corollary}
\newtheorem{prop}[thm]{Proposition}

\newtheorem{conj}{Conjecture}

\newtheorem{definition}{Definition}

\newtheorem{example}{Example}

\newcommand{\pmeas}[1]{P_{#1}}
\newcommand{\pdf}[1]{f_{#1}}
\newcommand{\pdfg}[1]{g_{#1}}

\newcommand{\qRV}[2]{[#1]_{#2}}
\newcommand{\qRVm}[1]{\qRV{#1}{m}}

\newcommand{\infodim}[1]{d(#1)}
\newcommand{\infodimu}[1]{\bar{d}(#1)}
\newcommand{\infodiml}[1]{\underline{d}(#1)}
\newcommand{\infodimlu}[1]{\overline{\underline{d}}(#1)}

\newcommand{\dimrate}{d'(\{X_t\})}
\newcommand{\dimratel}{\underline{d}'(\{X_t\})}
\newcommand{\dimratelu}{\overline{\underline{d}}'(\{X_t\})}
\newcommand{\dimrateu}{\overline{d}'(\{X_t\})}

\newcommand{\dimR}[1]{\mathsf{dim}_R(#1)}
\newcommand{\dimRu}[1]{\overline{\mathsf{dim}}_R(#1)}
\newcommand{\dimRl}[1]{\underline{\mathsf{dim}}_R(#1)}
\newcommand{\dimRlu}[1]{\overline{\underline{\mathsf{dim}}}_R(#1)}

\newcommand{\bfmu}{{\boldsymbol{\mu}}}
\newcommand{\covmat}[1]{C_{#1}}

\newcommand{\cov}[1]{K_{#1}}
\newcommand{\pcov}[1]{\overline{K}_{#1}}
\newcommand{\covt}[1]{K_{#1}(\tau)}
\newcommand{\pcovt}[1]{\overline{K}_{#1}(\tau)}

\newcommand{\psd}[1]{\mathsf{S}_{#1}}
\newcommand{\psdt}[1]{\mathsf{S}_{#1}(\theta)}
\newcommand{\sdf}[1]{\mathsf{F}_{#1}}
\newcommand{\psdf}[1]{\overline{\mathsf{F}}_{#1}}
\newcommand{\sdft}[1]{\mathsf{F}_{#1}(\theta)}
\newcommand{\sdfdt}[1]{\mathsf{F}'_{#1}(\theta)}
\newcommand{\dsdft}[1]{\mathsf{F}'_{#1}(\theta)}
\newcommand{\dsdf}[1]{\mathsf{F}'_{#1}}

\newcommand{\ent}[1]{H(#1)}
\newcommand{\diffent}[1]{h(#1)}
\newcommand{\mutinf}[1]{I(#1)}
\newcommand{\entrate}[1]{H'(#1)}
\newcommand{\kld}[2]{D(#1\|#2)}

\newcommand{\Xb}{\mathbf{X}}
\newcommand{\Yb}{\mathbf{Y}}

\newcommand{\yb}{\mathbf{y}}

\newcommand{\Zb}{\mathbf{Z}}
\newcommand{\Nb}{\mathbf{N}}
\newcommand{\Wb}{\mathbf{W}}
\newcommand{\Ub}{\mathbf{U}}
\newcommand{\Eb}{\mathbf{E}}

\newcommand{\reals}{\mathbb{R}}
\newcommand{\integers}{\mathbb{Z}}
\newcommand{\lipschitz}{\mathsf{K}}
\newcommand{\lebesgue}{\lambda}

\DeclareMathOperator*{\argmin}{arg\,min}
\newcommand{\expec}[1]{\mathsf{E}\left[#1\right]}
\newcommand{\Prob}{\mathbb{P}}
\newcommand{\diff}{\mathrm{d}}
\newcommand{\I}[1]{\mathbf{1}\!\left\{#1\right\}}
\newcommand{\e}[1]{e^{#1}}
\newcommand{\dom}[1]{\mathcal{#1}}

\newcommand{\limminf}{\varliminf_{m\to\infty}}
\newcommand{\limmsup}{\varlimsup_{m\to\infty}}
\newcommand{\limk}{\lim_{k\to\infty}}
\newcommand{\limm}{\lim_{m\to\infty}}

\newcommand{\trans}[1]{#1^{\textnormal{\textsf{\tiny T}}}} 
\newcommand{\herm}[1]{{#1}^\mathsf{H}}
\renewcommand{\conj}[1]{#1^*}
\newcommand{\rank}[1]{\mathrm{rank}(#1)}
\newcommand{\mat}[1]{#1}

\title{On the Information Dimension\\ of Stochastic Processes}

\author{Bernhard C. Geiger,~\IEEEmembership{Senior Member,~IEEE,} and Tobias Koch,~\IEEEmembership{Senior Member,~IEEE}
\thanks{The work of Bernhard C. Geiger has partly been funded by the Erwin Schr\"odinger Fellowship J 3765 of the Austrian Science Fund and by the German Ministry of Education and Research in the framework of an Alexander von Humboldt Professorship. The Know-Center is funded within the Austrian COMET Program - Competence Centers for Excellent Technologies - under the auspices of the Austrian Federal Ministry of Transport, Innovation and Technology, the Austrian Federal Ministry of Digital and Economic Affairs, and by the State of Styria. COMET is managed by the Austrian Research Promotion Agency FFG. The work of Tobias Koch has received funding from the European Research Council (ERC) under the European Union's Horizon 2020 research and innovation programme (grant agreement number 714161), from the 7th European Union Framework Programme under Grant 333680, from the Ministerio de Econom\'ia y Competitividad of Spain under Grants TEC2013-41718-R, RYC-2014-16332, and TEC2016-78434-C3-3-R (AEI/FEDER, EU), and from the Comunidad de Madrid under Grant S2103/ICE-2845. This work has been presented in part at the 2017 IEEE International Symposium on Information Theory, Aachen, Germany, June 2017, and at the 2018 International Zurich Seminar on Information and Communication, Zurich, Switzerland, February 2018.}
\thanks{Bernhard C. Geiger is with Know-Center GmbH, 8010, Graz, Austria (\mbox{e-mail:} \mbox{geiger@ieee.org}).}
\thanks{Tobias Koch is with the Signal Theory and Communications Department, Universidad Carlos III de Madrid, 28911, Legan\'es, Spain and also with the Gregorio Mara\~n\'on Health Research Institute, 28007, Madrid, Spain (\mbox{e-mail:} \mbox{koch@tsc.uc3m.es}).}
\thanks{Copyright (c) 2019 IEEE. Personal use of this material is permitted.  However, permission to use this material for any other purposes must be obtained from the IEEE by sending a request to \mbox{pubs-permissions@ieee.org}.}
}

\begin{document}

\maketitle

\begin{abstract}
In 1959, R\'enyi proposed the information dimension and the $d$-dimensional entropy to measure the information content of general random variables. This paper proposes a generalization of information dimension to stochastic processes by defining the information dimension rate as the entropy rate of the uniformly-quantized stochastic process divided by minus the logarithm of the quantizer step size $1/m$ in the limit as $m\to\infty$. It is demonstrated that the information dimension rate coincides with the rate-distortion dimension, defined as twice the rate-distortion function $R(D)$ of the stochastic process divided by $-\log(D)$ in the limit as $D\downarrow 0$. It is further shown that, among all multivariate stationary processes with a given (matrix-valued) spectral distribution function (SDF), the Gaussian process has the largest information dimension rate, and that the information dimension rate of multivariate stationary Gaussian processes is given by the average rank of the derivative of the SDF. The presented results reveal that the fundamental limits of almost zero-distortion recovery via compressible signal pursuit and almost lossless analog compression are different in general.
\end{abstract}


\section{Introduction}\label{sec:intro}
\IEEEPARstart{I}{n} 1959, R\'enyi \cite{Renyi_InfoDim} proposed the \emph{information dimension} and the \emph{$d$-dimensional entropy} to measure the information content of general random variables (RVs). His idea was to quantize the RV $X$ by a uniform quantizer of step size $1/m$, and to then analyze the entropy of the quantized RV $\qRVm{X}$ in the limit as $m$ tends to infinity. Assuming that the entropy $H(\qRVm{X})$ exists and the asymptotic expansion
\begin{equation}
H(\qRVm{X}) = d(X) \log m + H_{d}(X) + o(1)
\end{equation}
holds for $m\to\infty$ (where $o(1)$ refers to remainder terms that vanish as $m\to\infty$), R\'enyi referred to  $d(X)$ as the information dimension and to $H_d(X)$ as the $d$-dimensional entropy.

In recent years, it was shown that the information dimension is of relevance in various areas of information theory, including rate-distortion theory, almost lossless analog compression, or the analysis of interference channels. For example, Kawabata and Dembo \cite{Kawabata_RDDim} showed that the information dimension of a RV is equal to its \emph{rate-distortion dimension}, defined as twice the rate-distortion function $R(D)$ divided by $-\log(D)$ in the limit as $D\downarrow 0$. Koch \cite{Koch_SLB} demonstrated that the rate-distortion function of a source with infinite information dimension is infinite, and that for any source with finite information dimension and finite differential entropy the Shannon lower bound on the rate-distortion function is asymptotically tight. Wu and Verd\'u \cite{Wu_Renyi} analyzed linear encoding and Lipschitz decoding of discrete-time, independent and identically distributed (i.i.d.), stochastic processes and showed that the information dimension plays a fundamental role in achievability and converse results. Wu \emph{et al.} \cite{Wu_IF} showed that the degrees of freedom of the $K$-user Gaussian interference channel can be characterized through the sum of information dimensions. Stotz and B\"olcskei \cite{Stotz_IF} generalized this result to vector interference channels. 

Jalali and Poor \cite{Jalali_DimRate} proposed a generalization of information dimension to stationary, discrete-time, stochastic processes by defining the information dimension $\dimrate$ of the stochastic process $\{X_t\}$ as the information dimension of $(X_1,\ldots,X_k)$ divided by $k$ in the limit as $k\to\infty$.\footnote{More precisely, Jalali and Poor define the information dimension of a stochastic process via a conditional entropy of the uniformly-quantized process. For stationary processes, their definition coincides with the above-mentioned definition \cite[Lemma~3]{Jalali_DimRate}.} They showed that, for $\psi^*$-mixing processes, the information dimension is an achievable rate for universal compressed sensing with linear encoding and decoding via Lagrangian minimum entropy pursuit \cite[Th.~8]{Jalali_DimRate}. Rezagah \emph{et al.} \cite{Jalali_CompComp} showed that $\dimrate$ coincides, under certain conditions, with the rate-distortion dimension $\dimR{\{X_t\}}$, thus generalizing the result by Kawabata and Dembo~\cite{Kawabata_RDDim} to stochastic processes. Other notions of information dimensions for stochastic processes are discussed in \cite{gutman18_arxiv}.

In this paper, we propose a different definition for the information dimension of stationary, discrete-time, stochastic processes. Specifically, let $\{\qRVm{X_t}\}$ denote the stochastic process $\{X_t\}$ uniformly quantized with step size $1/m$. We define the \emph{information dimension rate} $\infodim{\{X_t\}}$ of $\{X_t\}$ as the entropy rate of $\{\qRVm{X_t}\}$ divided by $\log m$ in the limit as $m\to\infty$. For i.i.d.\ processes, our definition coincides with that of Jalali and Poor (and, in fact, evaluates to R\'enyi's information dimension of the marginal RV $X_t$). More generally, we show that these definitions are equivalent for \mbox{$\psi^*$-mixing} processes. Nevertheless, there are stochastic processes for which the two definitions disagree. In particular, we derive a closed-form expression for the information dimension rate of stationary, multivariate, Gaussian processes with \emph{power spectral density} (PSD) $\psd{X}$, which specialized to the univariate case yields that $\infodim{\{X_t\}}$ is equal to the Lebesgue measure of the set of harmonics on $[-1/2,1/2]$ where $\psd{X}$ is positive. For Gaussian processes with a bandlimited PSD, this implies that the information dimension rate $\infodim{\{X_t\}}$ is equal to twice the PSD's bandwidth. This is consistent with the intuition that for such processes not all samples contain information. For example, if the bandwidth of the PSD is $1/4$, then we expect that half of the samples in $\{X_t\}$ can be expressed as linear combinations of the other samples and, hence, do not contain information. In contrast, we show that the information dimension $\dimrate$ is $1$ if $\psd{X}$ is positive on any set with positive Lebesgue measure. In other words, $\dimrate$ does not capture the dependence of the information dimension on the support size of $\psd{X}$.

By emulating the proof of \cite[Lemma~3.2]{Kawabata_RDDim}, we further show that, for any stochastic process $\{X_t\}$, the information dimension rate $\infodim{\{X_t\}}$ coincides with the rate-distortion dimension $\dimR{\{X_t\}}$. This implies that $\dimrate$ coincides with $\dimR{\{X_t\}}$ only for those stochastic processes for which $\dimrate=\infodim{\{X_t\}}$.

The rest of this paper is organized as follows. In Section~\ref{sec:prelim}, we introduce the notation used in this paper. In Section~\ref{sec:RID}, we present preliminary results on the R\'enyi information dimension of RVs and random vectors. In Section~\ref{sec:dimrate}, we present our definition of the information dimension rate of a stochastic process, discuss its connection to the rate-distortion dimension, and compute the information dimension rate of stationary Gaussian processes. In Section~\ref{sec:poorrate}, we review the information dimension proposed by Jalali and Poor and discuss its relation to $\infodim{\{X_t\}}$. In Section~\ref{sec:operational}, we briefly discuss the operational meanings of information dimension in compressed sensing and zero-distortion recovery. Section~\ref{sec:discussion} concludes the paper with a discussion of the obtained results. Some of the proofs are deferred to the appendices.

\section{Notation and Preliminaries}\label{sec:prelim}

We denote by $\reals$, $\mathbb{C}$, and $\mathbb{Z}$ the set of real numbers, the set of complex numbers, and the set of integers, respectively. We further denote by $\mathbb{R}^+$ and $\mathbb{N}$ the set of nonnegative real numbers and the set of positive integers, respectively. We use a calligraphic font, such as $\dom{F}$, to denote other sets, and we denote complements as $\mathcal{F}^\mathsf{c}$. The set difference between two sets $\dom{F}$ and $\dom{G}$ is written as $\dom{F}\setminus\dom{G}$.

The real and imaginary parts of a complex number $z$ are denoted as $\mathfrak{Re}(z)$ and $\mathfrak{Im}(z)$, respectively, i.e., $z=\mathfrak{Re}(z)+\imath\mathfrak{Im}(z)$ where $\imath\triangleq \sqrt{-1}$. The complex conjugate of $z$ is denoted as $z^*$.

We use uppercase letters to denote deterministic matrices and boldface lowercase letters to denote deterministic vectors. The transpose of a vector or matrix is denoted by $\trans{(\cdot)}$, the Hermitian transpose by $
(\cdot)^\mathsf{H}$. The determinant and rank of a matrix $\mat{A}$ are $\det \mat{A}$ and $\rank{\mat{A}}$, respectively. We denote by $I_L$ the $L\times L$ identity matrix.

We denote RVs by uppercase letters, e.g., $X$. For a finite or countably infinite collection of RVs we abbreviate $X_{\ell}^k\triangleq (X_{\ell},\dots,X_{k-1},X_k)$, $X_\ell^\infty \triangleq (X_{\ell},X_{\ell+1},\dots)$, and $X_{-\infty}^k \triangleq (\dots,X_{k-1},X_k)$.\footnote{If $k<\ell$, then $X_\ell^k$ is the empty set.} Random vectors are denoted by boldface uppercase letters, e.g., $\Xb\triangleq\trans{(X_1,\dots,X_L)}$. Univariate discrete-time stochastic processes are denoted as $\{X_t,\,t\in\mathbb{Z}\}$ or, in short, as $\{X_t\}$. For $L$-variate stochastic processes we use the same notation but with $X_t$ replaced by $\Xb_t\triangleq\trans{(X_{1,t},\dots,X_{L,t})}$. We call $\{X_{i,t},\,t\in\mathbb{Z}\}$ a \emph{component process}. 

We denote the probability measure of the RV $X$ by $\pmeas{X}$. If $\pmeas{X}$ is absolutely continuous with respect to (w.r.t.) the Lebesgue measure, then we denote its probability density function (PDF) as $\pdf{X}$. We denote by $X_G$ a Gaussian RV with the same mean and variance as $X$, and we denote the corresponding Gaussian density as $\pdfg{X}$.

We define the quantization of a real-valued RV $X$ with precision $m$ as
\begin{equation}\label{eq:quantization}
 \qRV{X}{m} \triangleq \frac{\lfloor mX\rfloor}{m}
\end{equation}
where $\lfloor a\rfloor$ is the largest integer less than or equal to $a$. Likewise, $\lceil a \rceil$ denotes the smallest integer greater than or equal to $a$. We denote by $\qRV{X_{\ell}^k}{m} = (\qRV{X_{\ell}}{m},\ldots,\qRV{X_k}{m})$ the component-wise quantization of $X_{\ell}^k$ (and similarly for other finite or countably infinite collections of RVs and random vectors). For complex RVs $Z$ with real part $R$ and imaginary part $I$, the quantization $\qRV{Z}{m}$ is equal to $\qRV{R}{m}+\imath \qRV{I}{m}$. We define $\dom{C}(z_1^k,a)\triangleq[z_1,z_1+a) \times\cdots\times[z_k,z_k+a)$ as the \mbox{$k$-dimensional} hypercube in $\reals^k$, with its bottom-left corner at $z_1^k$ and with sidelength $a$. For example, we have that $\qRVm{X_1^k}=z_1^k$ if $X_1^k\in\dom{C}(z_1^k,1/m)$.

Let $H(\cdot)$, $h(\cdot)$, and $\kld{\cdot}{\cdot}$ denote entropy, differential entropy, and relative entropy, respectively, and let $\mutinf{\cdot;\cdot}$ denote mutual information~\cite{Cover_Information1}. We take logarithms to base \mbox{$e\approx 2.718$}, so mutual informations and entropies have dimension \emph{nats}. The entropy rate of a discrete-valued, stationary, $L$-variate process $\{\Xb_t\}$ is~\cite[Sec.~4.2]{Cover_Information1}
\begin{equation}
\label{eq:entrate}
\entrate{\{\Xb_t\}} \triangleq \lim_{k\to\infty} \frac{H(\Xb_1^k)}{k}.
\end{equation}
Note that the stationarity of $\{\Xb_t\}$ guarantees that the limit in \eqref{eq:entrate} exists and is equal to \cite[Th.~4.2.1]{Cover_Information1}
\begin{equation}
\label{eq:entrate_alt}
\lim_{k\to\infty} \frac{H(\Xb_1^k)}{k} = \lim_{k\to\infty} H(\Xb_1|\Xb^0_{-k}).
\end{equation}

We say that a stationary process $\{\Xb_t\}$ is \emph{$\psi^*$-mixing} if
\begin{equation}
 \limk \sup_{A,B}\frac{\pmeas{\Xb^0_{-\infty},\Xb_k^{\infty}}(A\cap B)}{\pmeas{\Xb^0_{-\infty}}(A)\pmeas{\Xb_k^{\infty}}(B)}=1
\end{equation}
where the supremum is over all $A\in\dom{F}^0_{-\infty}$ and $B\in\dom{F}_{k}^{\infty}$ satisfying $\pmeas{\Xb^0_{-\infty}}(A)\pmeas{\Xb_k^{\infty}}(B)>0$, and where $\dom{F}^0_{-\infty}$ and $\dom{F}_{k}^{\infty}$ are the $\sigma$-fields generated by $\Xb^0_{-\infty}$ and $\Xb_k^{\infty}$, respectively. The $\psi^*$-mixing property implies that $\{\Xb_t\}$ is \emph{information regular}, i.e., \cite[pp.~111-112]{Bradley_Mixing}
\begin{equation}\label{eq:inforegular}
 \limk \mutinf{\Xb_k^{\infty};\Xb^0_{-\infty}} = 0.
\end{equation}

\section{R\'enyi Information Dimension}
\label{sec:RID}
The \emph{R\'enyi information dimension} of a collection of RVs $X_1^k$ is defined as~\cite{Renyi_InfoDim}
\begin{equation}\label{eq:infodim}
 \infodim{X_1^k} \triangleq \limm \frac{\ent{\qRV{X_1^k}{m}}}{\log m}, \quad \textnormal{if the limit exists.}
\end{equation}
When the limit does not exist, we say that the information dimension does not exist. In this case, one may replace the limit either by the \emph{limit superior} or by the \emph{limit inferior} (denoted as $\varlimsup$ and $\varliminf$, respectively)
\begin{subequations}
\begin{IEEEeqnarray}{lCl}
 \infodimu{X_1^k} & \triangleq & \varlimsup_{m\to\infty} \frac{\ent{\qRV{X_1^k}{m}}}{\log m} \label{eq:infodimu} \\
 \infodiml{X_1^k} & \triangleq & \varliminf_{m\to\infty} \frac{\ent{\qRV{X_1^k}{m}}}{\log m} \label{eq:infodiml} 
\end{IEEEeqnarray}
\end{subequations}
and call $\infodimu{X_1^k}$ and $\infodiml{X_1^k}$ the upper and lower information dimension of $X_1^k$, respectively. Clearly,
\begin{equation}
\infodimu{X_1^k}= \infodiml{X_1^k}= \infodim{X_1^k}
\end{equation}
if the limit in \eqref{eq:infodim} exists.

We shall follow this notation throughout the document. Specifically, when reporting results in connection with limits, an overline $\overline{(\cdot)}$ indicates that the quantity in the brackets has been computed using the limit superior, an underline $\underline{(\cdot)}$ indicates that it has been computed using the limit inferior, both an overline and an underline $\overline{\underline{(\cdot)}}$ indicates that a result holds irrespective of whether the limit superior or limit inferior is taken. We write no lines if the limit exists.

\begin{definition}
For two RVs $X$ and $W$ with joint probability measure $\pmeas{X,W}$, the \emph{conditional information dimension} is defined as
\begin{equation}\label{eq:condinfodim}
 \infodim{X|W} \triangleq \limm \frac{\ent{\qRV{X}{m}|W}}{\log m} 
\end{equation}
provided the limit exists. If the limit does not exist, then we define the upper and lower conditional information dimension $\infodimu{X|W}$ and $\infodiml{X|W}$ by replacing the limit with the limit superior and the limit inferior, respectively.
\end{definition}

\subsection{Properties of Information Dimension}

The information dimension of a collection $X_1^k$ is bounded by the number of RVs in the collection, given the integer part of this collection has finite entropy.

\begin{lem}[{\cite[eq.~(7)]{Renyi_InfoDim},~\cite[Prop.~1]{Wu_Renyi}}]\label{lem:dim:finite}
  Let $X_1^k$ be a collection of real-valued RVs. If $\ent{\qRV{X_1^k}{1}}<\infty$, then
 \begin{equation}\label{eq:dim:bound}
 0\le \infodiml{X_1^k} \le \infodimu{X_1^k}\le k.
\end{equation}
If $\ent{\qRV{X_1^k}{1}}=\infty$, then $\infodimlu{X_1^k}=\infty$.
\end{lem}

Trivially, if $X_1^k$ is a collection of discrete RVs satisfying $\ent{\qRV{X_1^k}{1}}<\infty$, then $\infodim{X_1^k}= 0$. Moreover, if the joint distribution of $X_1^k$ is absolutely continuous w.r.t.\ the Lebesgue measure on $\reals^{k}$ and if $\ent{\qRV{X_1^k}{1}}<\infty$, then $\infodim{X_1^k}= k$~\cite[Th.~4]{Renyi_InfoDim}. More generally, R\'enyi claims that the information dimension of $X_1^k$ equals $n<k$ if the joint distribution of $X_1^k$ is absolutely continuous on some sufficiently smooth \mbox{$n$-dimensional} manifold in $\reals^{k}$~\cite[p.~209]{Renyi_InfoDim}. Furthermore, if $X$ is a real-valued RV satisfying $\ent{\qRV{X}{1}}<\infty$ and with probability measure
\begin{equation}
\label{eq:Renyi_mix}
\pmeas{X} = (1-\rho) \pmeas{d} + \rho \pmeas{c}
\end{equation}
where $\pmeas{d}$ is a discrete measure, $\pmeas{c}$ is an absolutely-continuous measure, and $0\leq\rho\leq 1$, then \cite[Th.~3]{Renyi_InfoDim}
\begin{equation}
\label{eq:ID_dc}
\infodim{X} = \rho.
\end{equation}

Two well-known properties of entropy are that it is reduced by conditioning \cite[Th.~2.6.5]{Cover_Information1} and that it obeys a chain rule. Furthermore, the conditional entropy of $X$ given $Y$ can be computed by first calculating the entropy conditioned on the event that $Y=y$, and by then averaging over $Y$. The corresponding results for information dimension are presented in the following three lemmas.

\begin{lem}\label{lem:dim:cond}
Suppose that $\ent{\qRV{X}{1}}<\infty$. Then, we have for any two RVs $X$ and $Y$
\begin{multline}
 \label{eq:lem_cond}
  \int \infodiml{X|Y=y} \diff\pmeas{Y}(y) \le \infodiml{X|Y}\\
  \le \infodimu{X|Y} \le \int \infodimu{X|Y=y} \diff\pmeas{Y}(y).
 \end{multline}
Consequently, if $\infodim{X|Y=y}$ exists $\pmeas{Y}$-almost surely, then the limit in~\eqref{eq:condinfodim} exists and 
  \begin{equation}
     \infodim{X|Y} =  \int \infodim{X|Y=y} \diff\pmeas{Y}(y).
  \end{equation}
\end{lem}
\begin{IEEEproof}
See Appendix~\ref{app:lem_cond}.
\end{IEEEproof}

\begin{lem}\label{lem:dim:conditioning_reduces}
For any two RVs $X$ and $Y$, we have
\begin{equation}
\infodimlu{X|Y} \leq \infodimlu{X}
\end{equation}
with equality if $X$ and $Y$ are independent.
\end{lem}
\begin{IEEEproof}
Since conditioning reduces entropy, we have $\ent{\qRV{X}{m}|Y}\leq\ent{\qRV{X}{m}}$, with equality if $X$ and $Y$ are independent. The lemma follows by dividing both sides of the inequality by $\log m$ and taking limits as $m\to\infty$.
\end{IEEEproof}

 \begin{lem}\label{lem:dim:chainrule}
 For the collection of RVs $X_1^k$, we have
 \begin{equation}
 \label{eq:dim:chainrule}
  \sum_{t=1}^k \infodimu{X_t} \ge \infodimu{X_1^k} \ge \infodiml{X_1^k} \ge \sum_{t=1}^k \infodiml{X_t|X_1^{t-1}}.
 \end{equation}
 \end{lem}
 
 \begin{IEEEproof}
 See Appendix~\ref{app:dim:chainrule}.
 \end{IEEEproof}

The left-most inequality in \eqref{eq:dim:chainrule} holds with equality if all information dimensions exist and the RVs $X_1^k$ are independent. There are examples where the right-most inequality is strict.
\begin{example}\label{ex:strict}
 Let $(X_1,X_2)$ be uniformly distributed on $[0,1]^2$ and let $Y=g(X_1,X_2)$, where $g{:}\ [0,1]^2\to[0,1]$ is bijective. Such a function can be constructed (see also the discussion in~\cite[Section~IV.B]{Wu_Renyi}). Since $g$ is bijective, we have $\infodim{Y|X_1,X_2}=\infodim{X_1,X_2|Y}=0$. Moreover, since $(X_1,X_2)$ is uniformly distributed on $[0,1]^2$, we have $\infodim{X_1,X_2}=2$. Finally, we have $\infodim{Y}\le 1$ by Lemma~\ref{lem:dim:finite}. From Lemma~\ref{lem:dim:chainrule}, we get
 \begin{equation}
 \infodim{X_1,X_2,Y}\ge \infodim{X_1,X_2} + \infodim{Y|X_1,X_2} = 2.
\end{equation}
 However, we also have
\begin{equation}
\infodim{Y} + \infodim{X_1,X_2|Y} \leq 1.
\end{equation}
It follows that
\begin{equation}
\infodim{X_1,X_2,Y} > \infodim{Y} + \infodim{X_1,X_2|Y}
\end{equation}
so the chain rule $\infodim{X_1,X_2,Y} \ge \infodim{Y} + \infodim{X_1,X_2|Y}$ holds with strict inequality.
\end{example}

The above example not only demonstrates that the chain rule for information dimension may hold with strict inequality, it also shows that the order in which the chain rule is expanded can be crucial.

\subsection{Information Dimension of Finite-Variance RVs}
For RVs $X_1^k$ that have a finite variance, the upper bound on $\infodimu{X_1^k}$ presented in Lemma~\ref{lem:dim:finite} can be tightened. To this end, we introduce further notation. We denote the covariance matrix of the vector $\Xb=\trans{(X_1,\dots,X_k)}$ by $\covmat{X_1^k}$. Furthermore, the cross-covariance matrix between $\Xb=\trans{(X_1,\dots,X_k)}$ and $\Yb=\trans{(Y_1,\dots,Y_k)}$ is denoted by $\covmat{X_1^kY_1^k}$, and the covariance matrix of the vector $\trans{(X_1,\ldots,X_k,Y_1,\ldots,Y_k)}$ is denoted by $\covmat{X_1^k,Y_1^k}$. Clearly,
\begin{equation}
 \covmat{X_1^k,Y_1^k} 
=\left[\begin{array}{cc}
           \covmat{X_1^k} & \covmat{X_1^kY_1^k} \\ \trans{\covmat{X_1^kY_1^k}} & \covmat{Y_1^k}
       \end{array}
\right].
\end{equation}

One can show that the information dimension of a collection of real-valued RVs $X_1^k$ cannot exceed the rank of its covariance matrix, i.e.,
\begin{equation}
\infodimlu{X_1^k} \le \rank{\covmat{X_1^k}}.
\end{equation}
This agrees with the intuition that linearly-dependent components of $X_1^k$ do not contribute to the information dimension. One can further show that collections of Gaussian RVs achieve this upper bound with equality. Thus, among all RVs with a given covariance structure, the Gaussian RV maximizes information dimension. These results follow directly from the more general results for stochastic processes (Theorem~\ref{thm:rate:Gauss}) in Section~\ref{sec:dimrate}.

The next theorem evaluates the conditional information dimension of $X_1^k$ given $Y_1^\ell$ for jointly Gaussian RVs $(X_1^k,Y_1^\ell)$.

\begin{thm}\label{thm:dim:condGauss}
Let $(X_1^k,Y_1^{\ell})$ be a collection of real-valued, jointly Gaussian RVs. The conditional information dimension of $X_1^k$ given $Y_1^{\ell}$ is equal to \begin{equation}
  \infodim{X_1^k|Y_1^\ell} = \rank{\covmat{X_1^k|Y_1^{\ell}}}
 \end{equation}
 where $\covmat{X_1^k|Y_1^{\ell}}$ is the generalized Schur complement of $\covmat{Y_1^{\ell}}$ in $\covmat{X_1^k,Y_1^{\ell}}$.
\end{thm}
\begin{IEEEproof}
See Appendix~\ref{proofs:dim:chainrule}.
\end{IEEEproof}
Theorem~\ref{thm:dim:condGauss} implies that the chain rule in Lemma~\ref{lem:dim:chainrule} holds with equality for Gaussian RVs. Indeed, if $X_1^k$ is a collection of real-valued, jointly Gaussian RVs, then we have $\infodim{X_1^k}=\rank{\covmat{X_1^k}}$ and $\infodim{X_1^\ell}=\rank{\covmat{X_1^\ell}}$. Moreover, by Theorem~\ref{thm:dim:condGauss}, $\infodim{X_{\ell+1}^k|X_1^\ell}$ equals the rank of the generalized Schur complement of $\covmat{X_1^\ell}$ in $\covmat{X_1^k}$, denoted by $\covmat{X_{\ell+1}^k|X_1^{\ell}}$. Since the rank of $\covmat{X_1^k}$ can be written as the sum of the ranks of $\covmat{X_1^\ell}$ and $\covmat{X_{\ell+1}^k|X_1^\ell}$~\cite[7.1.P28]{Horn_Matrix}, the claim follows.

\section{The Information Dimension Rate}\label{sec:dimrate}

We next propose the \emph{information dimension rate} as a generalization of information dimension to stochastic processes. We define the information dimension rate for general (possibly non-stationary) processes. However, for the sake of simplicity, most of our results will then be presented for stationary processes.

\begin{definition}\label{def:rate:infodim}
The \emph{information dimension rate} of the \mbox{$L$-variate} stochastic process $\{\Xb_t\}$ is defined as
 \begin{equation}
 \label{eq:dimpinrate}
   \infodim{\{\Xb_t\}} \triangleq \limm \limk \frac{\ent{\qRV{\Xb_1^k}{m}}}{k\log m}
 \end{equation}
 provided the limits exist. If the limits do not exist, then we define the upper and lower information dimension rate $\infodimu{\{\Xb_t\}}$ and $\infodiml{\{\Xb_t\}}$ by replacing the limits with the limits superior and limits inferior, respectively.
\end{definition}

\subsection{Properties of the Information Dimension Rate}

The information dimension rate satisfies properties similar to those presented in Lemma~\ref{lem:dim:finite} for the information dimension. We summarize them in the following lemma.

\begin{lem}\label{lem:rate:bounds}
 Let $\{\Xb_t\}$ be a stationary, $L$-variate, real-valued process. If $\ent{\qRV{\Xb_1}{1}}<\infty$, then
 \begin{equation}
 \label{eq:bounds}
   0 \le \overline{\underline{d}}(\{\Xb_t\}) \le \overline{\underline{d}}(\Xb_1) \le L.
    \end{equation}
If $\ent{\qRV{\Xb_1}{1}}=\infty$, then $\overline{\underline{d}}(\{\Xb_t\}) =\infty$.
\end{lem}

\begin{IEEEproof}
 Suppose first that $\ent{\qRV{\Xb_1}{1}}<\infty$. Then, the rightmost inequality in \eqref{eq:bounds} follows from~\eqref{eq:dim:bound}. The left-most inequality follows from the nonnegativity of entropy. Finally, the center inequality follows since conditioning reduces entropy, hence $\entrate{\{\qRVm{\Xb_t}\}} \le \ent{\qRVm{\Xb_1}}$.
 
 Now suppose that $\ent{\qRV{\Xb_1}{1}}=\infty$. Since $\qRV{\Xb_1}{1}$ is a function of $\qRVm{\Xb_1^k}$ for every $m,k \in\mathbb{N}$, we have 
\begin{equation}
 \ent{\qRV{\Xb_1}{1}} \le \ent{\qRV{\Xb_1^k}{m}}.
\end{equation}
This implies that $\entrate{\{\qRVm{\Xb_t}\}}=\infty$, and the claim that $\overline{\underline{d}}(\{\Xb_t\}) =\infty$ follows from Definition~\ref{def:rate:infodim}.
\end{IEEEproof}

The next result discusses how Lipschitz transformations affect the information dimension rate.

\begin{lem}\label{lem:rate:lipschitz}
 Let $\{\Xb_t\}$ be a stationary, $L$-variate, real-valued process, and let $\{f_t, t\in\integers\}$ be a sequence of Lipschitz functions from $\reals^L$ to $\reals^M$ with Lipschitz constants $\lipschitz_t$ satisfying
 \begin{equation}
 \sup_{t\in\integers} \lipschitz_t<\infty.
 \end{equation}
 Then,
\begin{equation}
\label{eq:rate:lipschitz}
  \infodimlu{\{f_t(\Xb_t)\}}\le  \infodimlu{\{\Xb_t\}}=\infodimlu{\{\Xb_t,f_t(\Xb_t)\}}.
\end{equation}
\end{lem}
\begin{IEEEproof}
See Appendix~\ref{proofs:dim:lipschitz}.
\end{IEEEproof}

If $\{f_t\}$ is a sequence of bi-Lipschitz functions with uniformly-bounded Lipschitz constants, then Lemma~\ref{lem:rate:lipschitz} implies that $\infodimlu{\{f_t(\Xb_t)\}}=  \infodimlu{\{\Xb_t\}}$. As a corollary, we thus obtain that the information dimension rate is invariant under scaling and translation. More generally, it follows that, if $\{\mathbf{c}_t\}$ and $\{W_t\}$ are sequences of $L$-variate vectors and $(L\times L)$-dimensional matrices, the latter satisfying $\sup_{t\in\integers} \|W_t\| < \infty$ and $\sup_{t\in\integers} \|W_t^{-1}\|< \infty$ for some induced matrix norm $\|\cdot\|$, then 
\begin{equation}
\infodimlu{\{W_t\Xb_t+\mathbf{c}_t\}} = \infodimlu{\{\Xb_t\}}.
\end{equation}
Since the information dimension rate of an i.i.d.\ process equals the information dimension of its marginal RVs, we further recover the well-known result that the information dimension of collections of RVs is invariant under scaling and translation~\cite[Lemma~3]{Wu_PhD}.

The next lemma shows that the information dimension rate of a collection of stochastic processes is unaffected by those that have zero information dimension rate.

\begin{lem}\label{lem:rate:mixture}
 Let $\{\Xb_t\}$ and $\{\Zb_t\}$ be two jointly stationary, $L$-variate, real-valued processes, and assume that $\infodim{\{\Zb_t\}}=0$. Then,
   \begin{IEEEeqnarray}{lCl} 
 \infodimlu{\{\Xb_t\},\{\Zb_t\}} & = & \infodimlu{\{\Xb_t\}} \label{eq:rate:mixture:eq1}\\
 \infodimlu{\{\Xb_t+\Zb_t\}} & = & \infodimlu{\{\Xb_t\}}. \label{eq:rate:mixture:eq3}
 \end{IEEEeqnarray}
 Moreover, if $Z$ is discrete with $\ent{Z}<\infty$, then we further have
 \begin{equation} 
   \infodimlu{\{\Xb_t\}|Z} \triangleq \limm \limk \frac{\ent{\qRV{\Xb_1^k}{m}|Z}}{k\log m} = \infodimlu{\{\Xb_t\}}.\label{eq:rate:mixture:eq2}
 \end{equation}
\end{lem}
\begin{IEEEproof}
See Appendix~\ref{app:dim:mixture}.
\end{IEEEproof}

\emph{Inter alia}, Lemma~\ref{lem:rate:mixture} can be used to compute the information dimension rate of a countable mixture of stochastic processes. For example, specialized to i.i.d.\ processes, \eqref{eq:rate:mixture:eq2} together with Lemma~\ref{lem:dim:cond} recovers \eqref{eq:ID_dc} by choosing $X_1\sim \pmeas{d}$, $X_2\sim \pmeas{c}$, and $\pmeas{Z}(1)=1-\pmeas{Z}(2)=1-\rho$.

\subsection{Information Dimension Rate vs.\ Rate-Distortion Dimension}\label{sec:dimrate:RDD}

 Let $R(\Xb_1^k,D)$ denote the rate-distortion function of the source $\Xb_1^k$, i.e.,
 \begin{equation}
 \label{eq:R2(D)}
 R(\Xb_1^k,D) \triangleq \inf_{\substack{P_{\hat{\Xb}_1^k|\Xb_1^k}\colon \\ \mathsf{E}[\|\hat{\Xb}_1^k-\Xb_1^k\|_2^2]\leq D}}  \mutinf{\Xb_1^k;\hat{\Xb}_1^k}
 \end{equation}
 where the infimum is over all conditional distributions of $\hat{\Xb}_1^k$ given $\Xb_1^k$ such that 
 \begin{equation}
  \label{eq:D}
  \mathsf{E}[\|\hat{\Xb}_1^k-\Xb_1^k\|_2^2]\leq D
 \end{equation}
 and where $\|\cdot\|_2$ denotes the Euclidean norm. We have the following definition.

\begin{definition}\label{def:Rddim}
The \emph{rate-distortion dimension} of the $L$-variate stochastic process $\{\Xb_t\}$ is defined as
\begin{equation}\label{eq:Ezra}
 \dimR{\{\Xb_t\}} \triangleq 2\lim_{D\downarrow 0} \limk \frac{R(\Xb_1^k,kD)}{-k \log D}
\end{equation}
provided the limits over $D$ and $k$ exist. (When the process $\{\Xb_t\}$ is stationary, the limit over $k$ always exists \cite[Th.~9.8.1]{gallager68}.) If the limits do not exist, then we define the upper and lower rate-distortion dimension $\dimRu{\{\Xb_t\}}$ and $\dimRl{\{\Xb_t\}}$ by replacing the limits with the limits superior and limits inferior, respectively.
\end{definition}

Intuitively, the rate-distortion function
\begin{equation}
R(D) \triangleq \lim_{k\to\infty} \frac{R(\Xb_1^k,kD)}{k}
\end{equation}
corresponds to the minimum number of nats per source symbol required to compress a stationary and ergodic source $\{\Xb_t\}$ with a vector quantizer of average per-symbol distortion not exceeding $D$ \cite[Sec.~9.8]{gallager68}. The rate-distortion dimension characterizes the growth of $R(D)$ as $D$ vanishes. For example, for an i.i.d.\ Gaussian source with variance $\sigma^2$, we have \cite[Th.~13.3.2]{Cover_Information1}
\begin{equation}
R(D) = \frac{1}{2}\log\left(\frac{\sigma^2}{D}\right) \I{0\leq D\leq \sigma^2}
\end{equation}
where $\I{\cdot}$ denotes the indicator function. Observe that in this case $R(D)$ grows like $1/2\log(1/D)$ as $D\to 0$. The rate-distortion dimension corresponds to twice the pre-log factor of the rate-distortion function $R(D)$, which in this case is $1$.

In contrast, the information dimension rate characterizes the growth of the entropy rate $\entrate{\{\qRV{\Xb_t}{m}\}}$ as $m$ increases. This entropy rate, in turn, corresponds essentially to the number of nats per source symbol required to compress each symbol $\Xb_t$ of a stationary and ergodic source $\{\Xb_t\}$ with a uniform quantizer of step size $1/m$. Since a symbol-wise, uniform quantizer cannot outperform the best vector quantizer, it follows that the information dimension rate is lower-bounded by the rate-distortion dimension.

For RVs, Kawabata and Dembo showed that the rate-distortion dimension is actually equal to its information dimension \cite[Prop.~3.3]{Kawabata_RDDim}. Thus, a symbol-wise, uniform quantizer achieves the same information dimension as the best vector quantizer. The following theorem generalizes this result to stochastic processes.

\begin{thm}\label{thm:rate:rddim}
For any $L$-variate, real-valued process $\{\Xb_t\}$, 
\begin{equation}
\label{eq:thm_main1}
 \dimRlu{\{\Xb_t\}} = \infodimlu{\{\Xb_t\}}.
\end{equation}
\end{thm}
\begin{IEEEproof}
See Appendix~\ref{proof:rddim}.
\end{IEEEproof}

Note that Theorem~\ref{thm:rate:rddim} also holds for non-stationary processes.

\subsection{Information Dimension Rate of Finite-Variance Processes}\label{sec:dimrate:gauss}
Let $\{\Xb_t\}$ be a stationary, $L$-variate, real-valued process with mean vector $\bfmu$ and (matrix-valued) spectral distribution function (SDF) $\theta\mapsto \sdft{\Xb}$. Thus, $\sdf{\Xb}$ is a bounded, non-decreasing, and right-continuous function on $[-1/2,1/2]$ such that the autocovariance function
\begin{equation}
\covt{\Xb} \triangleq \mathsf{E}\bigl[(\Xb_{t+\tau}-\bfmu)\trans{(\Xb_t-\bfmu)}\bigr], \quad \tau\in\mathbb{Z}
\end{equation}
is given by the Lebesgue-Stieltjes integral \cite[(7.3), p.~141]{Wiener_StochasticProcesses}
\begin{equation}
\label{eq:lebesgue_stieltjes}
\covt{\Xb} = \int_{-1/2}^{1/2} \e{-\imath 2\pi\tau\theta} \diff \sdft{\Xb}, \quad \tau\in\mathbb{Z}.
\end{equation}
It follows that the $(i,j)$-th element of $\sdf{\Xb}$ is the cross SDF $\theta\mapsto\sdft{X_iX_j}$ of the component processes $\{X_{i,t}\}$ and $\{X_{j,t}\}$, i.e.,
\begin{equation}
\covt{X_iX_j} = \int_{-1/2}^{1/2} \e{-\imath 2\pi\tau\theta} \diff \sdft{X_iX_j}, \quad \tau\in\mathbb{Z}
\end{equation}
where
\begin{equation}
\covt{X_iX_j} \triangleq \expec{(X_{i,t+\tau}-\mu_i)(X_{j,t}-\mu_j)}, \quad \tau\in\mathbb{Z}
\end{equation}
denotes the cross-covariance function. It further follows that the diagonal elements of $\sdf{\Xb}$ are real and non-decreasing, and they satisfy $\sdf{X_i}(1/2)-\sdf{X_i}(-1/2)=\sigma_i^2$, where $\sigma_i$ denotes the standard deviation of $X_{i,t}$. It can be shown that $\theta\mapsto \sdft{\Xb}$ has a derivative almost everywhere, which has positive semi-definite, Hermitian values \cite[(7.4), p.~141]{Wiener_StochasticProcesses}. We shall denote the derivative of $\sdf{\Xb}$ by $\dsdf{\Xb}$. When $\sdf{\Xb}$ is absolutely continuous w.r.t.\ the Lebesgue measure, its derivative $\dsdf{\Xb}$ coincides with the PSD $\psd{\Xb}$ of $\{\Xb_t\}$.

The following theorem shows that, among all processes of a given SDF, the Gaussian process maximizes the information dimension rate. It further characterizes the information dimension rate of such processes in terms of the SDF.

\begin{thm}\label{thm:rate:Gauss}
 Let $\{\Xb_t\}$ be a stationary, $L$-variate, real-valued process with SDF $\sdf{\Xb}$. Then,
 \begin{equation}\label{eq:thm:mainline}
  \infodimlu{\{\Xb_t\}} \le \int_{-1/2}^{1/2}  \rank{\dsdft{\Xb}} \diff \theta
 \end{equation}
with equality if $\{\Xb_t\}$ is Gaussian.
\end{thm}

\begin{IEEEproof}
See Appendix~\ref{proofs:rate:Gauss}.
\end{IEEEproof}

In order to prove Theorem~\ref{thm:rate:Gauss}, we invoke Bussgang's theorem to obtain an expression for the SDF of a quantized Gaussian process $\{\qRVm{\Xb_t}\}$ as a function of the SDF of the original process $\{\Xb_t\}$. Since we believe that this result is interesting on its own, we present it below.

\begin{lem}\label{lem:thm:bussgang}
 Let $\{\Xb_t\}$ be a stationary, $L$-variate, real-valued, Gaussian process with mean vector $\bfmu$ and SDF $\sdf{\Xb}$. Then, the $(i,j)$-th entry of the SDF $\theta\mapsto\sdft{\qRVm{\Xb}}$ of $\{\qRVm{\Xb_t}\}$ satisfies
  \begin{equation}\label{eq:SDMSum}
  \sdft{\qRVm{X_i}\qRVm{X_j}} = (a_i + a_j - 1) \sdft{X_iX_j} + \sdft{N_iN_j}
 \end{equation}
 where $N_{i,t}\triangleq X_{i,t}-\qRVm{X_{i,t}}$
 and
\begin{equation}
\label{eq:bounda}
 a_i\triangleq\frac{1}{\sigma_i^2}\expec{(X_{i,t}-\mu_i)(\qRVm{X_{i,t}}-\expec{\qRVm{X_{i,t}}})}.
\end{equation}
(In \eqref{eq:bounda}, $\mu_i$ and $\sigma_i$ denote the mean and standard deviation of $X_{i,t}$.) For every $i=1,\dots,L$, we have
 \begin{equation}\label{eq:boundBussgang}
  |1-a_i| \le \frac{1}{m}\sqrt{\frac{2}{\pi\sigma_i^2}}
 \end{equation}
 and
 \begin{equation}\label{eq:boundNoise}
  \int_{-1/2}^{1/2} \diff\sdft{N_i} \le \frac{1}{m^2}.
 \end{equation}
Moreover, if all component processes have zero mean and unit variance, then $a_1=\ldots=a_L$ and
\begin{equation}\label{eq:SDMSumEqual}
 \sdft{\qRVm{\Xb}} = (2a_1-1)\sdft{\Xb} + \sdft{\Nb}.
\end{equation}
\end{lem}
\begin{IEEEproof}
See Appendix~\ref{app:bussgang}.
\end{IEEEproof}

As a corrolary to Theorem~\ref{thm:rate:Gauss}, we obtain that for univariate, stationary, Gaussian processes with PSD $\psd{X}$, the information dimension rate is equal to the Lebesgue measure of the set of harmonics on $[-1/2,1/2]$ where $\psd{X}$ is positive, i.e.,
\begin{equation}\label{eq:LebesgueSupport}
 d(\{X_t\}) = \lambda(\{\theta{:}\ \psdt{X}>0\})
\end{equation}
where $\lambda(\cdot)$ denotes the Lebesgue measure. As pointed out by one of the reviewers, \eqref{eq:LebesgueSupport} can also be obtained directly by using the equivalence of information dimension rate and rate-distortion dimension (Theorem~\ref{thm:rate:rddim}) together with the parametric representation of the rate-distortion function \cite[eqs.~(9.7.42) \& (9.7.43)]{gallager68}
\begin{IEEEeqnarray}{rCl}
R(D_{\beta}) & = & \frac{1}{2}\int_{\mathcal{B}_{\beta}} \log\left(\frac{\psdt{X}}{\beta}\right)\diff \theta \\
D_{\beta} & = & \int_{-1/2}^{1/2} \min\{\psdt{X},\beta\}\diff\theta
\end{IEEEeqnarray}
for $\beta>0$, where $\mathcal{B}_{\beta}\triangleq \{\theta\in[-1/2,1/2]\colon \psdt{X}>\beta\}$. Indeed, when $\lambda(\mathcal{B}_0)$ is zero, we have $\infodim{\{X_t\}}=0$ since in this case the process $\{X_t\}$ has zero variance and, hence, the entropy rate of the quantized process $\{\qRV{X_t}{m}\}$ is zero, too.
When $\lambda(\mathcal{B}_0)$ is strictly positive, the distortion $D_{\beta}$ can be bounded as
\begin{equation}
\beta \lambda(\mathcal{B}_{\beta}) \leq D_{\beta} \leq \beta \lambda(\mathcal{B}_0).
\end{equation}
It follows by the continuity of the Lebesgue measure that $D_{\beta}/\beta \to \lambda(\mathcal{B}_0)$ as $\beta\to 0$. Consequently, $D_{\beta}\to 0$ if, and only if, $\beta\to 0$ and the rate-distortion dimension can be written as
\begin{IEEEeqnarray}{lCl}
\dimR{\{X_t\}} & = & \lim_{\beta\downarrow 0} \frac{\int_{\mathcal{B}_{\beta}} \log\left(\frac{\psdt{X}}{\beta}\right)\diff \theta}{-\log D_{\beta}} \nonumber\\
& = & \lim_{\beta\downarrow 0} \left\{\lambda(\mathcal{B}_{\beta})+\frac{\int_{\mathcal{B}_{\beta}}\log\psdt{X} \diff \theta}{-\log\beta} \right\} \nonumber\\
& = & \lambda(\mathcal{B}_0) + \lim_{\beta\downarrow 0} \frac{\int_{\mathcal{B}_{\beta}}\log\psdt{X} \diff \theta}{-\log\beta}. \label{eq:short_proof_1}
\end{IEEEeqnarray}
By the continuity of the Lebesgue measure, for every $\varepsilon>0$ there exists a $\beta'\in(0,1)$ such that $\lambda(\mathcal{B}_{\beta'})\geq \lambda(\mathcal{B}_0)-\varepsilon$. Since $\mathcal{B}_{\beta} \subseteq \mathcal{B}_0$, it follows that
\begin{equation}
\label{eq:setminus}
\lambda(\mathcal{B}_{\beta}\setminus \mathcal{B}_{\beta'}) \leq \lambda(\mathcal{B}_0) - \lambda(\mathcal{B}_{\beta'}) \leq \varepsilon, \quad \beta< \beta'.
\end{equation}
Thus, for every $0<\beta<\beta'<1$,
\begin{IEEEeqnarray}{lCl}
\int_{\mathcal{B}_{\beta}}\log\psdt{X} \diff \theta & = & \int_{\mathcal{B}_{\beta}\setminus \mathcal{B}_{\beta'}}\log\psdt{X} \diff \theta + \int_{\mathcal{B}_{\beta'}}\log\psdt{X} \diff \theta \nonumber\\
& \geq & \lambda(\mathcal{B}_{\beta}\setminus\mathcal{B}_{\beta'})\log\beta + \lambda(\mathcal{B}_{\beta'})\log\beta' \nonumber\\
& \geq & \varepsilon \log\beta + \lambda(\mathcal{B}_{\beta'})\log\beta'.\label{eq:short_proof_2}
\end{IEEEeqnarray}
Dividing both sides of \eqref{eq:short_proof_2} by $-\log\beta$, and letting first $\beta$ and then $\varepsilon$ tend to zero, we obtain that the second term on the RHS of \eqref{eq:short_proof_1} is nonnegative. However, by assumption the process $\{X_t\}$ has finite variance, so its PSD $\psd{X}$ is integrable over $[-1/2,1/2]$. Consequently, using the inequality $\log x \leq x -1$ and the nonnegativity of $\psd{X}$, we obtain that
\begin{equation}
\label{eq:short_proof_3}
\int_{\mathcal{B}_{\beta}}\log\psdt{X} \diff \theta \leq \int_{-1/2}^{1/2} \psdt{X} \diff \theta-1.
\end{equation}
Dividing both sides of \eqref{eq:short_proof_3} by $-\log\beta$, and letting $\beta$ tend to zero, we obtain that the second term on the RHS of \eqref{eq:short_proof_1} is also nonpositive. We conclude that this term is zero, so \eqref{eq:LebesgueSupport} follows from \eqref{eq:short_proof_1} and Theorem~\ref{thm:rate:rddim}. 

We observe from Theorem~\ref{thm:rate:Gauss} that the information dimension rate of a Gaussian process $\{\Xb_t\}$ depends only on the derivative of its SDF $\sdf{\Xb}$, which coincides almost everywhere with the derivative of the absolutely-continuous part of $\sdf{\Xb}$. Indeed, any SDF $\sdf{\Xb}$ can be decomposed as \cite[(4.3), p.~124]{Wiener_StochasticProcesses}
\begin{equation}
\label{eq:lebesgue_decomp}
\sdft{\Xb} = \sdft{\Xb,a} + \sdft{\Xb,d} + \sdft{\Xb,s}
\end{equation}
where $\sdf{\Xb,a}$ is \emph{absolutely continuous} w.r.t.\ the Lebesgue measure, $\sdf{\Xb,d}$ is \emph{discrete}, and $\sdf{\Xb,s}$ is \emph{singular}. Furthermore, $\sdfdt{\Xb} = \sdfdt{\Xb,a}$ almost everywhere \cite[Sec.~4]{Wiener_StochasticProcesses}. Consequently, the information dimension rate of a Gaussian process depends only on the absolutely-continuous part of its SDF. By combining \eqref{eq:lebesgue_decomp} with Theorem~\ref{thm:rate:Gauss} and Lemma~\ref{lem:rate:mixture}, we can show that the same is true for non-Gaussian processes.

\begin{cor}\label{cor:rate:lebesgue}
Let $\{\Xb_t\}$ be a stationary, $L$-variate, real-valued process with SDF $\sdf{\Xb}$, and let $\{\Xb_{t,a}\}$ be a stationary, $L$-variate, real-valued process with SDF $\sdf{\Xb,a}$, where $\sdf{\Xb,a}$ is the absolutely-continuous part of $\sdf{\Xb}$, cf.~\eqref{eq:lebesgue_decomp}. Then
\begin{equation}
\infodimlu{\{\Xb_t\}} = \infodimlu{\{\Xb_{t,a}\}}.
\end{equation}
\end{cor}
\begin{IEEEproof}
Combining the decomposition~\eqref{eq:lebesgue_decomp} with the spectral representation of stationary processes~\cite[Sec.~4.11]{Priestley}, it can be shown that every stationary process can be written as
\begin{equation}
\Xb_t = \Xb_{t,a} + \Xb_{t,d} + \Xb_{t,s}, \quad t\in\integers
\end{equation}
where $\{\Xb_{t,a}\}$, $\{\Xb_{t,d}\}$, and $\{\Xb_{t,s}\}$ are stationary, mutually uncorrelated, stochastic processes with the respective SDFs $\sdf{\Xb,a}$, $\sdf{\Xb,d}$, and $\sdf{\Xb,s}$; see \cite[p.~758]{Priestley} and references therein. Since $\dsdf{\Xb,d}$ and $\dsdf{\Xb,s}$ are zero almost everywhere \cite[Sec.~4]{Wiener_StochasticProcesses}, we obtain from Theorem~\ref{thm:rate:Gauss} and the nonnegativity of the information dimension rate (Lemma~\ref{lem:rate:bounds}) that
\begin{equation}
\label{eq:lebesgue_1}
\infodim{\{\Xb_{t,d}\}} = \infodim{\{\Xb_{t,s}\}} = 0.
\end{equation}
Corollary~\ref{cor:rate:lebesgue} follows by applying Lemma~\ref{lem:rate:mixture} first together with \eqref{eq:lebesgue_1} to show that
\begin{equation}
\label{eq:lebesgue_2}
\infodim{\{\Xb_{t,d}+\Xb_{t,s}\}}=0
\end{equation}
and then together with \eqref{eq:lebesgue_2} to show that
\begin{equation}
\infodimlu{\{\Xb_{t,a}+\Xb_{t,d}+\Xb_{t,s}\}} = \infodimlu{\{\Xb_{t,a}\}}.
\end{equation}
\end{IEEEproof}

\subsection{Information Dimension Rate of Complex-Valued Processes}

So far, we have considered real-valued stochastic processes. However, every complex-valued RV can be written as a two-dimensional, real-valued, random vector, so the previous results directly generalize to the complex case. In particular, one can define the information dimension rate of the $L$-variate, complex-valued process $\{\Zb_t\}$ as the information dimension rate of the $(2L)$-variate, real-valued process $\{\hat{\Xb}_t\}$ that follows by stacking the real part of $\Zb_t$ on top of the imaginary part of $\Zb_t$.

Let $\{\Zb_t\}$ be a stationary, $L$-variate, complex-valued process with mean vector $\bfmu$ and matrix-valued SDF $\sdf{\Zb}$, i.e., 
\begin{equation}\label{eq:rate:defSDFmatrix}
\covt{\Zb}=\int_{-1/2}^{1/2} e^{-\imath 2 \pi \tau \theta}\diff\sdft{\Zb}, \quad \tau\in\mathbb{Z}
\end{equation}
where
\begin{equation}
\covt{\Zb}\triangleq\expec{(\Zb_{t+\tau}-\bfmu)\herm{(\Zb_t-\bfmu)}}, \quad \tau\in\mathbb{Z}
\end{equation}
is the autocovariance function. We say that a stationary, $L$-variate, complex-valued process $\{\Zb_t\}$ is \emph{proper} if it has finite variance, its mean vector is the zero vector, and its pseudo-autocovariance function satisfies
\begin{equation}\label{eq:rate:defPSDFmatrix}
 \pcovt{\Zb} \triangleq \expec{\Zb_{t+\tau}\trans{\Zb_{t}}} = 0, \quad \tau\in\integers.
\end{equation}
The following result generalizes Theorem~\ref{thm:rate:Gauss} to complex-valued stochastic processes.

\begin{thm}\label{thm:rate:proper}
Let $\{\Zb_t\}$ be a stationary, $L$-variate, complex-valued process with matrix-valued SDF $\sdf{\Zb}$. Then,
\begin{equation}\label{eq:upperbound}
 \infodimlu{\{\Zb_t\}} \le 2 \int_{-1/2}^{1/2}\rank{\dsdft{\Zb}} \diff\theta
\end{equation}
with equality if $\{\Zb_t\}$ is Gaussian and proper.
\end{thm}

\begin{IEEEproof}
 See Appendix~\ref{proofs:rate:proper}.
\end{IEEEproof}

Note that neither Gaussianity nor properness is sufficient for equality in Theorem~\ref{thm:rate:proper}. Conversely, Gaussianity and properness are not necessary for equality. For example, any univariate stationary Gaussian process achieves \eqref{eq:upperbound} with equality if its real and imaginary components are independent and if the derivatives of their SDFs have matching support.

\section{Another Definition of Information Dimension}\label{sec:poorrate}
Jalali and Poor \cite{Jalali_DimRate} proposed a different definition for the information dimension of a univariate stochastic process. We shall refer to this information dimension as the \emph{block-average information dimension} and denote it by $\dimrate$. In this section, we discuss scenarios in which the information dimension rate (Definition~\ref{def:rate:infodim}) coincides with and differs from the block-average information dimension. For ease of exposition, in this section we follow \cite{Jalali_DimRate} and restrict our attention to univariate real-valued processes.

The following definition for the information dimension of stochastic processes was proposed in \cite{Jalali_DimRate}.
\begin{definition}\label{def:averageblock}
The \emph{block-average information dimension} of the stochastic process $\{X_t\}$ is defined as
 \begin{equation}
 \label{eq:dimrate:block:orig}
  \dimrate \triangleq \limk \limm \frac{\ent{\qRV{X_k}{m}|\qRV{X_1^{k-1}}{m}}}{\log m} 
 \end{equation}
 provided the limits exist. If the limits do not exist, then one can define the upper and lower block-average information dimension $\dimrateu$ and $\dimratel$ by replacing the limits by limits superior and limits inferior, respectively.
\end{definition}

In the following, we restrict ourselves to stationary processes, in which case the limit over $k$ in \eqref{eq:dimrate:block:orig} is guaranteed to exist. We refer to $\dimrate$ as the block-average information dimension because it was shown in \cite[Lemma~3]{Jalali_DimRate} that, if $\{X_t\}$ is stationary and the information dimension $\infodim{X_1^k}$ exists for every $k$, then
\begin{equation}\label{eq:dimrate:block}
 \dimrate = \limk \frac{\infodim{X_1^k}}{k}.
\end{equation}
If $\infodim{X_1^k}$ does not exist, then the proof of~\cite[Lemma~3]{Jalali_DimRate} reveals that
\begin{equation}\label{eq:dimrate:relation}
 \dimratel \le \limk\frac{\infodiml{X_1^k}}{k} \le \limk\frac{\infodimu{X_1^k}}{k} = \dimrateu.
\end{equation}

Since conditioning reduces entropy, it follows immediately that
\begin{equation}
\label{eq:dimrateUB}
\dimratelu \leq \infodimlu{X_t}.
\end{equation}
Thus, like the information dimension rate, the block-average information dimension of the stochastic process $\{X_t\}$ cannot exceed the information dimension of the marginal RV $X_t$.

While the entropy rate $\entrate{\{X_t\}}$ of a stationary process $\{X_t\}$ can alternatively be written as the conditional entropy of $X_1$ given $X_{-\infty}^0$, cf.~\eqref{eq:entrate_alt}, the block-average information dimension $\dimrate$ does, in general, not permit a similar expression. In fact, let
\begin{equation}
\infodim{X_1|X_{-\infty}^0}\triangleq \lim_{m\to\infty}\lim_{k\to\infty} \frac{\ent{\qRVm{X_1}|X_{-k}^0}}{\log m}
\end{equation}
provided the limit over $m$ exists. (Since conditioning reduces entropy, the limit over $k$ always exists.) The upper and lower information dimensions $\infodimu{X_1|X_{-\infty}^0}$ and $\infodiml{X_1|X_{-\infty}^0}$ are defined analogously by replacing the limit over $m$ by the limit superior and limit inferior, respectively. 
Then, we have that
\begin{equation}\label{eq:comparison:cond}
\infodimlu{X_1|X_{-\infty}^0} \leq \dimratelu
\end{equation}
where the inequality can be strict; see Theorem~\ref{thm:dimrates} and Example~\ref{ex:Gauss_kicks_ass} below.

\subsection{Block-Average Information Dimension vs.\\ Information Dimension Rate}

We next demonstrate that, for $\psi^*$-mixing processes, the information dimension rate $\infodim{\{X_t\}}$ coincides with the block-average information dimension $\dimrate$. However, in general the two definitions do not coincide, but there exists an ordering between them.

\begin{thm}\label{thm:dimrates}
 Let $\{X_t\}$ be a stationary process. Then,
 \begin{equation}
 \label{eq:dimrates}
  \infodimlu{X_1|X_{-\infty}^0} \le \infodimlu{\{X_t\}} \le \dimratelu.
 \end{equation}
  Moreover,
  \begin{IEEEeqnarray}{lCl}
\IEEEeqnarraymulticol{3}{l}{\limk \varliminf_{m\to\infty} \frac{I(\qRV{X_k}{m};\qRV{X_{-\infty}^0}{m}|\qRV{X_1^{k-1}}{m})}{\log m}} \nonumber\\
\quad & \leq & \dimratelu - \infodimlu{\{X_t\}}\nonumber\\
& \leq & \limk \varlimsup_{m\to\infty} \frac{I(\qRV{X_k}{m};\qRV{X_{-\infty}^0}{m}|\qRV{X_1^{k-1}}{m})}{\log m} \label{eq:thm_dimrates}
  \end{IEEEeqnarray}
    where the limits over $k$ exist because, by the stationarity of $\{X_t\}$, the mutual information $I(\qRV{X_k}{m};\qRV{X_{-\infty}^0}{m}|\qRV{X_1^{k-1}}{m})$ is monotonically decreasing in $k$. 
\end{thm}

\begin{IEEEproof}
See Appendix~\ref{app:thm:dimrates}.
\end{IEEEproof}

The inequalities in \eqref{eq:thm_dimrates} imply that, if the limits over $m$ exist, then
\begin{equation}
\label{eq:thm_dimrates_2}
\limk \limm \frac{I(\qRV{X_k}{m};\qRV{X_{-\infty}^0}{m}|\qRV{X_1^{k-1}}{m})}{\log m} = 0
\end{equation}
is a necessary and sufficient condition for the equality of $\infodimlu{\{X_t\}}$ and $\dimratelu$. Note that, for every $m=2,3,\ldots$, we have \cite[eq.~(8.9)]{Gray_Entropy}
\begin{equation}
\limk I(\qRV{X_k}{m};\qRV{X_{-\infty}^0}{m}|\qRV{X_1^{k-1}}{m}) = 0.
\end{equation}
Thus, \eqref{eq:thm_dimrates_2} is satisfied for processes $\{X_t\}$ that allow us to change the order of taking limits as $k$ and $m$ tend to infinity. However, in general \eqref{eq:thm_dimrates_2} is difficult to check. We next present a sufficient condition that is easier to verify.

\begin{cor}
\label{cor:dimrates}
Let $\{X_t\}$ be a stationary process. Assume that there exists a nonnegative integer $n$ such that
\begin{equation}
\label{eq:dimRateCond}
\mutinf{X_1^k;X_{-\infty}^{-n}} < \infty, \quad k=1,2,\ldots
\end{equation}
Then, $\infodimlu{\{X_t\}}=\dimratelu$.
\end{cor}
\begin{IEEEproof}
See Appendix~\ref{proof:cor_dimrates}.
\end{IEEEproof}

Condition \eqref{eq:dimRateCond} holds for $\psi^*$-mixing processes. Indeed, since every $\psi^*$-mixing process satisfies \eqref{eq:inforegular}, it follows that one can find an $n$ such that $\mutinf{X_1^\infty;X_{-\infty}^{-n}}<\infty$. The condition \eqref{eq:dimRateCond} holds then by the data processing inequality.

If~\eqref{eq:dimRateCond} holds for $n=0$, then we even have that
\begin{equation}
\label{eq:boring}
\infodimlu{X_1|X_{-\infty}^0} =\infodimlu{\{X_t\}}= \dimratelu = \infodimlu{X_t}.
\end{equation}
Thus, in this case all presented generalizations of information dimension to stochastic processes coincide with the information dimension of the marginal RV. To prove \eqref{eq:boring}, we note that \eqref{eq:dimRateCond} with $n=0$ gives
\begin{equation}
\label{eq:boring_2}
 \mutinf{X_1;X_{-\infty}^0}<\infty.
\end{equation}
It then follows by the data processing inequality that
\begin{equation}
\mutinf{\qRV{X_1}{m};X_{-\infty}^0}\leq\mutinf{X_1;X_{-\infty}^0}<\infty.
\end{equation}
Consequently,
\begin{align}
 \infodim{X_1|X_{-\infty}^0} & = \limm \frac{\ent{\qRV{X_1}{m}}-\mutinf{\qRV{X_1}{m};X_{-\infty}^0}}{\log m}\nonumber\\
 & =\infodim{X_1}
\end{align}
if the limit exists. In general, we have $ \infodimlu{X_1|X_{-\infty}^0}=\infodimlu{X_t}$. The claim \eqref{eq:boring} follows then by \eqref{eq:dimrates} and because, by \eqref{eq:dimrateUB}, $\dimratelu\leq \infodimlu{X_t}$. 

Condition \eqref{eq:dimRateCond} with $n=0$ is satisfied, for example, if $\{X_t\}$ is a sequence of i.i.d.\ RVs, if it is a discrete-valued stochastic process with finite marginal entropy, or if it is a continuous-valued stochastic process with finite marginal differential entropy and finite differential entropy rate.

In the following, we present two examples of processes $\{X_t\}$ for which $\infodimlu{X_1|X_{-\infty}^0} =\infodimlu{\{X_t\}}= \dimratelu$. As we shall argue, neither of these examples satisfies \eqref{eq:dimRateCond}, hence \eqref{eq:dimRateCond} is sufficient but not necessary.

\begin{example}\label{ex:piecewise}
 Let $\{B_t\}$ be a sequence of i.i.d.\ Bernoulli-$\rho$ RVs, i.e., $\pmeas{B_t}(0)=1-\pmeas{B_t}(1)=\rho$, and let $\{Y_t\}$ be a sequence of i.i.d.\ RVs with PDF $\pdf{Y}$ supported on $[0,1]$ and finite differential entropy. By \eqref{eq:ID_dc}, we thus have that $\infodim{Y_t}=1$ for every $t$. We define the stochastic process $\{X_t\}$ as
\begin{equation}
\label{eq:piecewise_def}
X_t=B_t X_{t-1} + (1-B_t) Y_t, \quad t\in\mathbb{Z}
\end{equation}
and assume that $X_t$ has the same marginal distribution as $Y_t$. Note that $\{X_t\}$ is first-order Markov, so
\begin{equation}
\infodim{X_1|X_{-\infty}^0}=\infodim{X_1|X_0}=\rho\infodim{Y}=\rho.
\end{equation}
Furthermore, \cite[Th.~3]{Jalali_DimRate} demonstrates that $\dimrate=\rho$. Thus, together with \eqref{eq:dimrates}, this yields that
\begin{equation}
\infodim{X_1|X_{-\infty}^0}=\infodim{\{X_t\}}=\dimrate=\rho.
\end{equation}
The stochastic process $\{X_t\}$, as defined by \eqref{eq:piecewise_def}, satisfies \eqref{eq:thm_dimrates_2} but not \eqref{eq:dimRateCond}. Indeed, for every nonnegative integer $n$, we have $\mutinf{X_1;X_{-n}}=\infty$, since $X_1$ has finite differential entropy and the event $X_1=X_{-n}$ has positive probability. It follows that $I(X_1^k;X_{-\infty}^{-n})=\infty$ for every $k$ and $n$, so \eqref{eq:dimRateCond} is violated. In contrast, we have
 \begin{multline}
 \label{eq:piecewise}
 \varlimsup_{m\to\infty} \frac{I(\qRV{X_k}{m};\qRV{X_{-\infty}^0}{m}|\qRV{X_1^{k-1}}{m})}{\log m} \\
 = \varlimsup_{m\to\infty} \frac{I(\qRV{X_k}{m};\qRV{X_{-\infty}^0}{m}|\qRV{X_1^{k-1}}{m},B_k)}{\log m}
 \end{multline}
since conditioning on the binary random variable $B_k$ changes mutual information by at most one bit. If $B_k=1$, then $\qRV{X_k}{m}=\qRV{X_{k-1}}{m}$; if $B_k=0$, then $\qRV{X_k}{m}=\qRV{Y_k}{m}$, which is independent of $\qRV{X_{-\infty}^{k-1}}{m}$. In both cases, the conditional mutual information between $\qRV{X_k}{m}$ and $\qRV{X_{-\infty}^0}{m}$ given $\qRV{X_1^{k-1}}{m}$ is zero, so \eqref{eq:thm_dimrates_2} is satisfied.
\end{example}

\begin{example}\label{ex:the_other_example}
 Let the process $\{\tilde{X}_t\}$ be periodic with period $P\in\mathbb{N}$ and have finite marginal differential entropy. Further let $\Delta$ be uniformly distributed on $\{0,\dots,P-1\}$. Then, the shifted process $\{X_t\}$, defined by
 \begin{equation}
 \label{eq:the_other_example}
 X_t=\tilde{X}_{t+\Delta}, \quad t\in\integers
 \end{equation}
 is stationary \cite[Th.~10-5]{Papoulis_Probability} and has finite marginal differential entropy. For every $k=P,P+1,\ldots$ and $m=2,3,\ldots$, we have that $\ent{\qRV{X_1}{m}|X_{-k+1}^0}=0$ and $\ent{\qRV{X_1^k}{m}}=\ent{\qRV{X_1^P}{m}}$, hence
 \begin{equation}
 \infodim{X_1|X_{-\infty}^0}=\infodim{\{X_t\}}=\dimrate=0.
 \end{equation}
As in the previous example, the stochastic process $\{X_t\}$ satisfies \eqref{eq:thm_dimrates_2} but not \eqref{eq:dimRateCond}. Indeed, for every nonnegative integer $n$, we have $\mutinf{X_1^k;X_{-\infty}^{-n}}=\infty$ since $X_1$ has finite differential entropy and the process is periodic. In contrast, $\qRV{X_k}{m}=\qRV{X_{k-P}}{m}$, so the conditional mutual information between $\qRV{X_k}{m}$ and $\qRV{X_{-\infty}^0}{m}$ given $\qRV{X_1^{k-1}}{m}$ is zero when $k=P+1,P+2\ldots$
\end{example}

In many cases, the inequalities in Theorem~\ref{thm:dimrates} can be strict. The following example shows such a strict inequality for the class of stationary Gaussian processes $\{X_t\}$ with PSD supported on a set of positive Lebesgue measure.\footnote{The assumption that $\{X_t\}$ has a PSD is made for notational convenience and is not essential. All steps in Example~\ref{ex:Gauss_kicks_ass} continue to hold if we replace $\psd{X}$ by the derivative of the SDF $\sdf{X}$.}

\begin{example}
\label{ex:Gauss_kicks_ass}
Let $\{X_t\}$ be a stationary Gaussian process with zero mean, variance $\sigma^2$, and PSD $\psd{X}$ having support $\mathcal{B}_0$. It follows from Theorem~\ref{thm:rate:Gauss} that
 \begin{equation}
 \infodim{\{X_t\}} = \lambda\left(\mathcal{B}_0\right).
 \end{equation}
 We next argue that if $0<\lambda(\mathcal{B}_0)<1$ then $\dimrate=1$ and $\infodim{X_1|X_{-\infty}^0}=0$. Consequently,
 \begin{equation}
 \infodim{X_1|X_{-\infty}^0} < \infodim{\{X_t\}} < \dimrate.
 \end{equation}
To show that $\dimrate=1$, we note that
\begin{align}
\dimratel & = \limk \varliminf_{m\to\infty}\frac{\ent{\qRV{X_k}{m}|\qRV{X_1^{k-1}}{m}}}{k \log m} \nonumber\\
& \geq \limk \varliminf_{m\to\infty} \frac{\ent{\qRV{X_0}{m}|X_{-k}^{-1}}}{\log m} \label{eq:ex_ISIT1}
\end{align}
where the inequality follows by the stationarity of $\{X_t\}$; because conditioning reduces entropy; and because, conditioned on $X_{-k}^{-1}$, $\qRV{X_0}{m}$ is independent of $\qRV{X_{-k}^{-1}}{m}$. Since $\{X_t\}$ is Gaussian, it follows that, conditioned on $X_{-k}^{-1}$, the RV $X_0$ is Gaussian with mean $\mathsf{E}[X_0|X_{-k}^{-1}]$ and variance $\sigma_k^2$, which is independent of $X_{-k}^{-1}$. It can be further shown that if $\lambda(\mathcal{B}_0)>0$, then $\sigma_k^2>0$ for every finite $k$ (see Lemma~\ref{lem:TobiPredictionError} below). It follows that, conditioned on any $X_{-k}^{-1}=x_{-k}^{-1}$, the RV $X_0$ has a PDF, so by \eqref{eq:ID_dc}
\begin{equation}
\label{eq:ex_ISIT2}
\lim_{m\to\infty} \frac{\ent{\qRV{X_0}{m}\bigm|X_{-k}^{-1}=x_{-k}^{-1}}}{\log m} = 1, \quad k=1,2,\ldots
\end{equation}
Together with Fatou's lemma, this shows that the RHS of \eqref{eq:ex_ISIT1} is $1$, hence $\dimrate=1$.
 
 To demonstrate that $\infodim{X_1|X_{-\infty}^0}=0$, we note that \mbox{$\lambda(\mathcal{B}_0)<1$} implies that
 \begin{equation}
 \int_{-1/2}^{1/2} \log \psdt{X}\diff \theta = -\infty.
 \end{equation}
 This is a necessary and sufficient condition for $\sigma_k^2\to 0$ as $k\to\infty$; see, e.g., \cite[Sec.~10.6]{grenanderszego58}. Intuitively, the fact that $\sigma_k^2\to 0$ implies that the conditional distribution of $X_1$ given $X_{-\infty}^0$ is almost surely degenerate, hence $\infodim{X_1|X_{-\infty}^0}=0$. To prove this rigorously, we apply \cite[Lemma~30]{Wu_PhD} together with the fact that conditioning reduces entropy to upper-bound
 \begin{equation}
 \label{eq:UFF}
 \ent{\qRVm{X_1}|X_{-k}^0} \leq \ent{\qRVm{X_1-\mathsf{E}[X_1|X_{-k}^{0}]}} + \log 2.
 \end{equation}
 Expressing $X_1-\mathsf{E}[X_1|X_{-k}^{0}]$ as $\sigma_{k+1} Z$, where $Z$ is zero-mean, unit-variance Gaussian, the RHS of \eqref{eq:UFF} can be written as $\ent{\lfloor m\sigma_{k+1} Z\rfloor} + \log 2$. Since $\sigma_{k}^2\to 0$ as $k\to\infty$, we obtain from \cite[Lemma~1]{Koch_SLB} that
 \begin{equation}
 \lim_{k\to\infty} \ent{\qRVm{X_1}|X_{-k}^0} \leq \log 2, \quad m=2,3,\ldots
 \end{equation}
Consequently, the claim follows from the definition of $\infodim{X_1|X_{-\infty}^0}$.
 \end{example}

\begin{lem}\label{lem:TobiPredictionError}
Let $\{X_t\}$ be a stationary, univariate, real-valued, Gaussian process with zero mean, variance $\sigma^2$, and SDF $\sdf{X}$. Suppose that $\sigma_k^2=0$ for some finite $k$. Then,
 \begin{equation}
  \lambda(\{\theta : \sdfdt{X}>0\})=0.
 \end{equation}
\end{lem}
\begin{IEEEproof}
 See Appendix~\ref{proof:TobiPredictionError}.
\end{IEEEproof}

\subsection{Block-Average Information Dimension vs.\\ Rate-Distortion Dimension}\label{sec:poorrate:rddim}
The connection between the block-average information dimension and the rate-distortion dimension of a stochastic process was studied in \cite{Jalali_CompComp}. The equivalence between the rate-distortion dimension and the information dimension \cite[Prop.~3.3]{Kawabata_RDDim} directly implies that
\begin{IEEEeqnarray}{lCl}
 \dimrateu & = & 2\limk \varlimsup_{D\to 0} \frac{R(X_1^k,kD)}{-k \log D} \label{eq:dimrate_vs_infodim_1}
\end{IEEEeqnarray}
Rezagah \emph{et al.} \cite{Jalali_CompComp} demonstrated that the order of the limits on the RHS of \eqref{eq:dimrate_vs_infodim_1} can be exchanged. More precisely, \cite[Th.~2]{Jalali_CompComp} states that if $\lim_{D\to 0} \frac{R(X_1^k,kD)}{-k \log D}$ exists for all $k$, then
\begin{equation}\label{eq:rezagah}
  \dimR{\{X_t\}} = \dimrateu.
\end{equation}
This may appear as a contradiction to our results, since we demonstrate in Theorem~\ref{thm:rate:rddim} that $\dimRlu{\{X_t\}}=\infodimlu{\{X_t\}}$, and Example~\ref{ex:Gauss_kicks_ass} demonstrates that there are stochastic processes for which $\infodimlu{\{X_t\}}<\dimratelu$. However, the proof of \eqref{eq:rezagah} relies on the fact that \cite[Sec.~VI-E]{Jalali_CompComp}
\begin{IEEEeqnarray}{lCl}
\label{eq:rezagah_2}
 \left|\frac{1}{k} R(X_1^k,kD)- \lim_{\kappa\to\infty}\frac{1}{\kappa} R(X_1^{\kappa},\kappa D)\right| & \le & \frac{1}{k} \mutinf{X_1^k;X_{-\infty}^0} \IEEEeqnarraynumspace
\end{IEEEeqnarray}
and that the RHS of \eqref{eq:rezagah_2} vanishes as $k\to\infty$. If \eqref{eq:boring_2} holds, then this is indeed the case; see \cite[eqs.~(8.6)--(8.10)]{Gray_Entropy}. As shown in Corollary~\ref{cor:dimrates}, in this case we also have that $\dimratelu=\infodimlu{\{X_t\}}$. In fact, as discussed after Corollary~\ref{cor:dimrates}, in this case all presented generalizations of information dimension to stochastic processes coincide with the information dimension of the marginal RV. In contrast, if \eqref{eq:boring_2} does not hold then, by the data processing inequality, the RHS of \eqref{eq:rezagah_2} is infinite. This is, for example, the case if $\{X_t\}$ is a stationary process with positive variance and a PSD that is zero on a set of positive Lebesgue measure, since for such processes the differential entropy $h(X_1|X_{-\infty}^0)$ is $-\infty$. Our proof of Theorem~\ref{thm:rate:rddim} does not rely on \eqref{eq:rezagah_2}. We thus conclude that $\dimRlu{\{X_t\}}=\infodimlu{\{X_t\}}$ for every stochastic process $\{X_t\}$, but that $\dimRlu{\{X_t\}}=\dimratelu$ only for those processes for which $\dimratelu=\infodimlu{\{X_t\}}$.

\section{Operational Characterizations}
\label{sec:operational}
Information dimension was recently given an operational characterization in almost lossless data compression \cite{Wu_Renyi}. Specifically, Wu and Verd\'u defined the \emph{minimum $\epsilon$-achievable rate} $R(\epsilon)$ to be the minimum of $R>0$ such that there exists a sequence of encoders $f_k\colon \reals^k \to \reals^{\lfloor Rk\rfloor}$ and decoders $g_k\colon \reals^{\lfloor Rk\rfloor}\to \reals^k$ satisfying \cite[Def.~4]{Wu_Renyi}
\begin{equation}
\label{eq:lossless_recovery}
\Prob[g_k(f_k(X_1^k))\neq X_1^k] \leq \epsilon
\end{equation}
for all $k$ sufficiently large. As argued in \cite[Sec.~IV-B]{Wu_Renyi}, if we impose no restrictions on $f_k$ and $g_k$, then zero rate is achievable even for $\epsilon=0$, since the cardinality of $\reals^k$ is the same for any $k$. However, if we restrict ourselves either to encoders $f_k$ that are linear or to decoders $g_k$ that are Lipschitz continuous, then the minimum $\epsilon$-achievable rate for collections of i.i.d. RVs $X_1^k$ with a discrete-continuous mixed distribution, i.e., a distribution of the form \eqref{eq:Renyi_mix}, is given by
\begin{equation}
R(\epsilon) = \infodim{X}.
\end{equation}
Thus, for such RVs, information dimension has an operational characterization.

For stochastic processes $\{X_t\}$, Wu and Verd\'u further demonstrated that the minimum $\epsilon$-achievable rate, achievable with Lipschitz-continuous decoders $g_k$, can be lower-bounded as \cite[Remark~4]{Wu_Comp}
\begin{equation}
\label{eq:lipschitz_dec}
R(\epsilon) \geq \dimrateu - \epsilon.
\end{equation}
To the best of our knowledge, for non-i.i.d.\ processes $\{X_t\}$, no matching achievability result exists for almost lossless data compression.

In contrast, for universal compressed sensing with linear encoding and decoding via Lagrangian minimum entropy pursuit, it was shown by Jalali and Poor that $\dimrateu$ is an achievable rate when $\{X_t\}$ is 
$\psi^*$-mixing:
\begin{thm}[{\cite[Th.~8]{Jalali_DimRate}}]\label{thm:operational}
 Consider a $\psi^*$-mixing stationary process $\{X_t\}$ taking value in $[0,1]$ with upper block-average information dimension $\dimrateu$. For each $k$, let the entries of the measurement matrix $A\in\mathbb{R}^{\ell \times k}$ be drawn i.i.d.\ according to a zero-mean, unit-variance, Gaussian distribution. Given $X_1^k$ generated by $\{X_t\}$ and $\trans{(Y_1,\ldots,Y_{\ell})} = A \trans{(X_1,\ldots,X_k)}$, let
 \begin{equation}
  \hat{\Xb} \triangleq \argmin_{\mathbf{u}\in\dom{X}_m^k} \left\{\hat{H}_j(\mathbf{u})+\frac{\gamma}{k^2}\Vert A \mathbf{u}-\trans{(Y_1,\ldots,Y_{\ell})}\Vert_2^2\right\}
 \end{equation}
 where $\dom{X}_m\triangleq\{[x]_{2^m}\colon\ x\in[0,1]\}$, $\hat{H}_j(\cdot)$ is the conditional empirical entropy~\cite[Def.~1]{Jalali_DimRate}, $m=\lceil r \log \log k\rceil$ (for $r>1$), $j=o(\frac{\log k}{\log \log k})$, and $\gamma= (\log k)^{2r}$. If the number of measurements $\ell=\ell_k$ satisfies
 \begin{equation}
 \frac{\ell_k}{k}\ge (1+\delta)\dimrateu, \quad \textnormal{for some arbitrary $\delta>0$}
 \end{equation}
  then
 \begin{equation}
 \frac{1}{\sqrt{k}}\Vert\hat{\Xb}- \trans{(X_1,\ldots,X_k)}\Vert_2 \to 0 \quad \textnormal{in probability}
 \end{equation}
 as $k\to\infty$.
\end{thm}

In words, Theorem~\ref{thm:operational} states that if the rate of random linear measurements of $X_1^k$ is slightly larger than the block-average information dimension, then the Lagrangian relaxation of minimum entropy pursuit provides an asymptotically distortion-free estimate of $X_1^k$ in terms of the Euclidean norm. Thus, for $\psi^*$-mixing processes, the block-average information dimension is an achievable rate for almost zero-distortion recovery.

We next discuss an operational characterization of the rate-distortion dimension. By Theorem~\ref{thm:rate:rddim}, this is also an operational characterization of the information dimension rate. In \cite{Jalali_CompComp}, Rezagah \emph{et al.} considered the almost zero-distortion recovery of stationary processes when the decoder employs \emph{compressible signal pursuit (CSP)} optimization:

\begin{thm}[{\cite[Cor.~2]{Jalali_CompComp}}]
\label{cor:Jalali}
Consider a stationary, real-valued process $\{X_t\}$ and a system of random linear observations $\trans{(Y_1,\ldots,Y_{\ell})}=A \trans{(X_1,\ldots,X_k)}$ with measurement matrix $A\in\reals^{\ell\times k}$ composed of i.i.d.\ zero-mean, unit-variance, Gaussian RVs. If the number of measurements $\ell=\ell_k$ satisfies
\begin{equation}
\varliminf_{k\to\infty} \frac{\ell_k}{k} > \dimRu{\{X_t\}}
\end{equation}
then there exists a family of compression codes such that
\begin{equation}
 \frac{1}{\sqrt{k}}\Vert\hat{\Xb}- \trans{(X_1,\ldots,X_k)}\Vert_2 \to 0 \quad \textnormal{in probability}
 \end{equation}
 as $k\to\infty$, where $\hat{\Xb}$ is the solution of the CSP optimization
 \begin{equation}
 \hat{\Xb} = \argmin_{\mathbf{u}\in\dom{C}_{k}} \Vert\trans{(Y_1,\ldots,Y_{\ell})} - A \mathbf{u}\Vert_2
 \end{equation}
 and $\dom{C}_{k}$ denotes the codebook of the compression code.
\end{thm}

In words, if the rate of random linear measurements of $X_1^k$ is slightly larger than the rate-distortion dimension, then there exists a family of compression codes for which CSP optimization yields an asymptotically distortion-free estimate of $X_1^k$ in terms of the Euclidean norm. Thus, the rate-distortion dimension is an achievable rate for almost zero-distortion recovery.

To summarize, \eqref{eq:lipschitz_dec} demonstrates that $\dimrateu$ yields a lower bound on the sampling rate required for almost lossless recovery with Lipschitz-continuous decoders. In contrast, Theorem~\ref{cor:Jalali} demonstrates that $\dimRu{\{X_t\}}$ (and hence also $\infodimu{\{X_t\}}$) is an achievable rate for almost zero-distortion recovery. Furthermore, as illustrated by Example~\ref{ex:Gauss_kicks_ass}, there are processes $\{X_t\}$ for which
\begin{equation}
\label{eq:basta}
\infodim{\{X_t\}} = \dimR{\{X_t\}} < \dimrate.
\end{equation}
Our results thus demonstrate that there exist stationary processes for which the sampling rate required for almost zero-distortion recovery is strictly smaller than the sampling rate required for almost lossless recovery with Lipschitz-continuous decoders. In other words, the fundamental limits of almost zero-distortion recovery and almost lossless recovery are different in general.

Comparing the lower bound \eqref{eq:lipschitz_dec} for almost lossless recovery with Theorem~\ref{cor:Jalali} for almost zero-distortion recovery, we observe that there are two main differences in the setup:
\begin{itemize}
\item[i)] \eqref{eq:lipschitz_dec} is obtained for a Lipschitz-continuous decoder $g_k$, whereas Theorem~\ref{cor:Jalali} is based on CSP optimzation;
\item[ii)] for almost lossless recovery, $\hat{\Xb}=g_k(f_k(X_1^k))$ is required to be exactly equal to $X_1^k$ with high probability (cf.~\eqref{eq:lossless_recovery}), whereas for almost zero-distortion recovery it suffices that $\frac{1}{\sqrt{k}}\Vert\hat{\Xb}-
\trans{(X_1,\ldots,X_k)}\Vert_2$ be small.
\end{itemize}
The following example presents a class of stationary processes for which almost zero-distortion recovery at rate $\infodimu{\{X_t\}}$ may also be achieved with linear encoders and decoders. This suggests that the second difference has greater impact.

\begin{example}
Let $\{X_t\}$ be a stationary, univariate, real-valued, Gaussian process possessing a PSD $\psd{X}$ with support $[-1/4,1/4]$. By Theorem~\ref{thm:rate:Gauss}, we have that $\infodim{\{X_t\}}=1/2$. We next invoke the sampling theorem to demonstrate that there exist linear encoders $f_k\colon\reals^k\to\reals^{\ell_k}$ and decoders $g_k\colon\reals^{\ell_k}\to\reals^k$ such that
\begin{equation}
\lim_{k\to\infty} \frac{\ell_k}{k} = \frac{1}{2}
\end{equation}
and
\begin{equation}
\label{eq:ex8}
\frac{1}{\sqrt{k}}\Vert\hat{\Xb}-
\trans{(X_1,\ldots,X_k)}\Vert_2 \to 0 \quad \textnormal{in probability}
\end{equation}
as $k\to\infty$, where $\hat{\Xb}=g_k(f_k(X_1^k))$.

To describe $f_k$ and $g_k$, we divide the indices $t=1,\ldots,k$ into three groups:
\begin{IEEEeqnarray}{lCl}
\dom{I}_1 & \triangleq & \{1,\ldots,\Delta_k\} \cup \{k-\Delta_k+1,\ldots,k\} \\
\dom{I}_2 & \triangleq & \{2i\colon i\in\integers\} \cap \{\Delta_k + 1,\ldots,k-\Delta_k\} \\
\dom{I}_3 & \triangleq & \{2i+1 \colon i \in\integers\} \cap \{\Delta_k + 1,\ldots,k-\Delta_k\}
\end{IEEEeqnarray}
where $\{\Delta_k\}$ is an arbitrary sequence of even integers that tends to infinity sublinearly in $k$. The encoder $f_k$ only reproduces the values of $X_t$ with indices $t\in\dom{I}_1\cup \dom{I}_2$, i.e., $f_k(X_1^k) = \{X_t,\,t\in \dom{I}_1\cup \dom{I}_2\}$. Consequently,
\begin{equation}
\ell_k = 2\Delta_k + \left\lfloor\frac{k-2\Delta_k}{2}\right\rfloor
\end{equation}
and the rate $\ell_k/k$ converges to $\frac{1}{2}$ as $k\to\infty$.

We next show that we can find a decoder $g_k$ for which \eqref{eq:ex8} holds. Clearly, the values $\{X_t,\,t\in \dom{I}_1\cup \dom{I}_2\}$ are directly observed. It therefore remains to estimate the missing values of $X_1^k$, which is done via the interpolation formula 
\begin{equation}
\hat{X}_t = \sum_{i=-\Delta_k/2}^{\Delta_k/2} X_{2i+t-1} \frac{\sin\left(\pi\left[\frac{1}{2}-i\right]\right)}{\pi \left(\frac{1}{2}-i\right)}, \quad t\in\dom{I}_3.
\end{equation}
It follows that
\begin{IEEEeqnarray}{lCl}
\IEEEeqnarraymulticol{3}{l}{\expec{\Vert \trans{(\hat{X}_1,\ldots,\hat{X}_k)}-\trans{(X_1,\ldots,X_k)}\Vert_2^2}} \nonumber\\
\quad & = & \sum_{t\in\dom{I}_3} \expec{(\hat{X}_t-X_t)^2} \nonumber\\
& = & \left\lceil\frac{k-2\Delta_k}{2}\right\rceil \expec{(\hat{X}_{\Delta_k+1}-X_{\Delta_k+1})^2} \label{eq:ex8_2}
\end{IEEEeqnarray}
where the last step is due to stationarity. By the sampling theorem for stochastic processes, the expected value on the RHS of \eqref{eq:ex8_2} vanishes as $\Delta_k\to\infty$ \cite[Th.~1]{Balakrishnan}. Thus, dividing both sides of \eqref{eq:ex8_2} by $k$ and letting $k\to\infty$ gives
\begin{equation}
\lim_{k\to\infty}\frac{1}{k}\expec{\Vert \trans{(\hat{X}_1,\ldots,\hat{X}_k)}-\trans{(X_1,\ldots,X_k)}\Vert_2^2} = 0
\end{equation}
which together with Chebyshev's inequality \cite[Th.~4.10.7]{AsDo00} yields \eqref{eq:ex8}.
\end{example}

\section{Conclusions}\label{sec:discussion}
R\'enyi \cite{Renyi_InfoDim} proposed the information dimension and the $d$-dimensional entropy to measure the information content of general RVs. His idea was to quantize the real-valued RV $X$ by a uniform quantizer of step size $1/m$, and to then analyze the entropy of the quantized RV $\qRVm{X}$ in the limit as $m$ tends to infinity. His results demonstrate that any RV with positive information dimension has infinite information content. This is, e.g., the case for RVs whose probability measure has an absolutely-continuous part. The problem becomes even more interesting for stochastic processes $\{X_t\}$, since their information content is not only determined by the distribution of the marginals $X_t$, but also by their temporal dependence. For example, consider a stationary Gaussian process $\{X_t\}$ with bandlimited PSD. On the one hand, Gaussian processes have absolutely-continuous marginals, so one would expect that their information content is infinite. On the other hand, for processes with a bandlimited PSD, the present sample $X_0$ can be perfectly predicted from its infinite past $X_{-1},X_{-2},\ldots$ (see Example~\ref{ex:Gauss_kicks_ass}), which suggests that the information content of $\{X_t\}$ is zero.

To shed some light on such questions, we proposed a generalization of information dimension to stochastic processes by defining the information dimension rate $\infodim{\{\Xb_t\}}$ as the entropy rate $\entrate{\{\qRV{\Xb_t}{m}\}}$ divided by $\log m$ in the limit as $m\to\infty$. We demonstrated that the information dimension rate coincides with the rate-distortion dimension, defined as twice the pre-log factor of the rate-distortion function $R(D)$. We further showed that among all stationary process with PSD $\psd{\Xb}$, the Gaussian process has the largest information dimension rate. This is consistent with the observation that Gaussian processes are the hardest to predict, hence they are expected to have the largest information content. We then showed that the information dimension rate of stationary Gaussian processes is given by the average rank of $\psd{\Xb}$, i.e.,
\begin{equation}
\infodim{\{\Xb_t\}} = \int_{-1/2}^{1/2} \rank{\psdt{\Xb}}  \diff \theta.
\end{equation}
Specialized to the univariate case, this yields that the information dimension rate is given by the Lebesgue measure of the support of $\psd{X}$, i.e.,
\begin{equation}
\infodim{\{X_t\}}=\lambda(\{\theta : \psdt{X}>0\}).
\end{equation}
This agrees with the intuition that if the PSD of $\{X_t\}$ is zero on a set of positive Lebesgue measure, then some samples can be expressed in terms of the remaining samples and have therefore no information content. It further answers the above question whether stationary Gaussian processes with a bandlimited PSD have infinite information content in the positive, unless the PSD is zero almost everywhere. 

An alternative definition for the information dimension of a stochastic process was proposed by Jalali and Poor \cite{Jalali_DimRate} as the information dimension of $X_1^k$ divided by $k$ in the limit as $k\to\infty$. We referred to this quantity as the block-average information dimension $\dimrate$. While $\infodim{\{X_t\}}$ and $\dimrate$ coincide for $\psi^*$-mixing processes, in general we have that $\infodim{\{X_t\}}\leq\dimrate$, where the inequality can be strict. In particular, as illustrated by Example~\ref{ex:Gauss_kicks_ass}, if the support of $\psd{X}$ of the Gaussian process $\{X_t\}$ has positive Lebesgue measure, then $\dimrate=1$. Thus, in contrast to the information dimension rate, the block-average information dimension does not capture the dependence of the information dimension on the support size of $\psd{X}$.

The essential difference between the definitions of $\infodim{\{X_t\}}$ and $\dimrate$ is the order in which the limits over the quantization bin size $1/m$ and the block size $k$ are taken. Rezagah \emph{et al.} \cite{Jalali_CompComp} showed that these limits can be exchanged if the process satisfies
\begin{equation}\label{eq:finiteRedundancyRate}
I(X_1;X_{-\infty}^0)< \infty
\end{equation}
in which case $\dimR{\{X_t\}}=\dimrate$. However, in this case the information dimension of the stochastic process $\{X_t\}$ coincides with the information dimension of the marginal RV $X_t$. In other words, for such processes a generalization of information dimension to stochastic processes is redundant. In contrast, we showed in Theorem~\ref{thm:rate:rddim} that, for any stochastic process $\{X_t\}$, the information dimension rate $\infodim{\{X_t\}}$ coincides with the rate-distortion dimension $\dimR{\{X_t\}}$. This implies that $\dimrate$ coincides with $\dimR{\{X_t\}}$ only for those stochastic processes for which $\dimrate=\infodim{\{X_t\}}$.

The equivalence between the information dimension rate $\infodim{\{X_t\}}$ and the rate-distortion dimension $\dimR{\{X_t\}}$ implies that $\infodim{\{X_t\}}$ inherits the operational characterizations of $\dimR{\{X_t\}}$. For example, it was demonstrated in \cite{Jalali_CompComp} that $\dimR{\{X_t\}}$ is an achievable rate for almost zero-distortion recovery. In contrast, \cite{Wu_Comp} shows that $\dimrate$ is a lower bound on the minimum $\epsilon$-achievable rate, achievable with Lipschitz-continuous decoders. By demonstrating that there are processes for which
\begin{equation}
\infodim{\{X_t\}} = \dimR{\{X_t\}} < \dimrate
\end{equation}
our results show that the fundamental limits of almost zero-distortion recovery and almost lossless recovery are different in general. Jalali and Poor \cite{Jalali_DimRate} further showed that $\dimrate$ is an achievable rate for universal lossless compressed sensing with linear encoding and decoding via Lagrangian minimum entropy pursuit when $\{X_t\}$ is $\psi^*$-mixing. Since for $\psi^*$-mixing processes we have $\infodim{\{X_t\}}=\dimrate$, our definition also inherits this operational characterization.

\appendices

\section{Appendix to Section~\ref{sec:RID}}

\subsection{Proof of Lemma~\ref{lem:dim:cond}}
\label{app:lem_cond}
The first inequality in \eqref{eq:lem_cond}, namely,
 \begin{multline}
   \int \limminf \frac{\ent{\qRV{X}{m}|Y=y}}{\log m} \diff\pmeas{Y}(y)\\
  \le \limminf \int\frac{\ent{\qRV{X}{m}|Y=y}}{\log m}\diff\pmeas{Y}(y)
 \end{multline}
follows directly from Fatou's lemma \cite[Th.~1.6.8, p.~50]{AsDo00}. The second inequality follows because the limit inferior is upper-bounded by the limit superior. For the third inequality, note that for every $m=2,3,\ldots$ and $Y=y$ \cite[eq.~(11)]{Renyi_InfoDim}
\begin{equation}\label{eq:upperboundreverseFatou1}
\frac{\ent{\qRV{X}{m}|Y=y}}{\log m} \leq \frac{\ent{\qRV{X}{1}|Y=y}}{\log 2} + 1.
\end{equation}
Furthermore, since conditioning reduces entropy, we have
\begin{IEEEeqnarray}{lCl}
\int \left(\frac{\ent{\qRV{X}{1}|Y=y}}{\log 2} + 1\right) \diff\pmeas{Y}(y) & = & \frac{\ent{\qRV{X}{1}|Y}}{\log 2} + 1 \nonumber\\
& \leq & \frac{\ent{\qRV{X}{1}}}{\log 2} + 1 \nonumber\\
& < & \infty
\end{IEEEeqnarray}
for every $m=2,3,\ldots$ Hence, the RHS of~\eqref{eq:upperboundreverseFatou1} is integrable, and the third inequality in \eqref{eq:lem_cond} follows again by Fatou's lemma.

\subsection{Proof of Lemma~\ref{lem:dim:chainrule}}
\label{app:dim:chainrule}
 If $\ent{\qRV{X_1^k}{1}}=\infty$, then we have $\infodim{X_1^k}=\infty$ and the right-most inequality in \eqref{eq:dim:chainrule} holds trivially. Moreover, in this case $\ent{\qRV{X_t}{1}}=\infty$ for at least one $t$, so for this $t$ we also have $\infodimu{X_t}=\infty$. Thus, also the left-most inequality holds.
 
If $\ent{\qRV{X_1^k}{1}}<\infty$, then we have
 \begin{equation}
 \ent{\qRV{X_t}{1}|X_1^{t-1}}\le \ent{\qRV{X_t}{1}} < \infty, \quad t=1,\ldots,k
 \end{equation}
 hence the upper information dimensions are finite. It follows by the chain rule of entropy and because conditioning reduces entropy that
 \begin{align}
 \infodimu{X_1^k} & = \limmsup \sum_{t=1}^k \frac{\ent{\qRV{X_t}{m}|\qRV{X_1^{t-1}}{m}}}{\log m}\nonumber\\
& \leq \sum_{t=1}^k \limmsup \frac{\ent{\qRV{X_t}{m}}}{\log m} \nonumber\\
& = \sum_{i=t}^k \infodimu{X_t}.
 \end{align}
 Likewise, we have
 \begin{align}
   \infodiml{X_1^k} &= \limminf \sum_{t=1}^k \frac{\ent{\qRV{X_t}{m}|\qRV{X_1^{t-1}}{m}}}{\log m}\nonumber\\
   &\ge  \sum_{t=1}^k \limminf \frac{\ent{\qRV{X_t}{m}|X_1^{t-1}}}{\log m}\nonumber\\
   &= \sum_{t=1}^k \infodiml{X_t|X_1^{t-1}}
 \end{align}
 where the inequality follows because conditioning reduces entropy and because, conditioned on $X_1^{t-1}$, $\qRV{X_t}{m}$ is independent of $\qRV{X_1^{t-1}}{m}$.

\subsection{Proof of Theorem~\ref{thm:dim:condGauss}}\label{proofs:dim:chainrule}
To simplify notation, we shall write collections of RVs as vectors, namely, $\Xb=\trans{(X_1,\ldots,X_k)}$ and $\Yb=\trans{(Y_1,\ldots,Y_{\ell})}$. The proof of Theorem~\ref{thm:dim:condGauss} is based on the following lemma.

\begin{lem}\label{lem:GaussRVError}
 Let $\Xb$ and $\Yb$ be $k$- and $\ell$-dimensional, jointly Gaussian vectors with mean vectors $\bfmu_{\Xb}$ and $\bfmu_{\Yb}$ and joint covariance matrix $\covmat{\Xb,\Yb}$. Then, there exists a $k\times \ell$ matrix $T$ and a length-$k$ vector $\bfmu$ such that $\expec{\Xb|\Yb} = \bfmu + T\Yb$. Moreover, $\Eb\triangleq\Xb-\bfmu-T\Yb$ has zero mean, is uncorrelated with $\Yb$, and satisfies $\covmat{\Eb} = \covmat{\Xb} - T\covmat{\Yb}\trans{T}$.
\end{lem}
\begin{IEEEproof}
If $\Xb$ and $\Yb$ are jointly Gaussian, then $\Xb$ can be written as a linear transformation of $\Yb$ and an uncorrelated error. This follows from the fact that, for jointly Gaussian $\Xb$ and $\Yb$, the linear minimum mean-square error (LMMSE) estimator of $\Xb$ given $\Yb$ always exists and is given by $\expec{\Xb|\Yb} = \bfmu + T\Yb$. The result that $\Eb$ has zero mean, is uncorrelated with $\Yb$, and satisfies $\covmat{\Eb} = \covmat{\Xb} - T\covmat{\Yb}\trans{T}$ follows by direct calculation.
\end{IEEEproof}

Since information dimension is translation invariant, it follows that
\begin{IEEEeqnarray}{lCl}
 \infodim{\Xb|\Yb=\yb} & = & \infodim{\bfmu+T\Yb+\Eb|\Yb=\yb} \nonumber\\
 & = & \infodim{\Eb|\Yb=\yb}. \label{eq:dim:chainrule:1}
\end{IEEEeqnarray}
Furthermore, since $\Xb$ and $\Yb$ are jointly Gaussian, so are $\Yb$ and $\Eb$, and from the fact that they are uncorrelated follows that they are independent. Thus,
\begin{equation}
\label{eq:dim:chainrule:2}
 \infodim{\Eb|\Yb=\yb} =  \infodim{\Eb} = \rank{\covmat{\Eb}}
\end{equation}
where $\covmat{\Eb}$ is the covariance matrix of $\Eb$. The identities \eqref{eq:dim:chainrule:1} and \eqref{eq:dim:chainrule:2} hold for every $\yb$, so it follows from Lemma~\ref{lem:dim:cond} that $\infodim{\Xb|\Yb}=\rank{\covmat{\Eb}}$. It remains to show that $\covmat{\Eb}$ is the generalized Schur complement of $\covmat{\Yb}$ in $\covmat{\Xb,\Yb}$. Indeed, by~\cite[7.1.P28]{Horn_Matrix} there exists a matrix $W$ such that $\covmat{\Yb\Xb} = \covmat{\Yb}W$. The generalized Schur complement of $\covmat{\Yb}$ in $\covmat{\Xb,\Yb}$ is then given by
\begin{equation}
\label{eq:dim:chainrule:3}
\covmat{\Xb|\Yb} = \covmat{\Xb} - \herm{W}\covmat{\Yb} W.
\end{equation}
Comparing \eqref{eq:dim:chainrule:3} with the expression of $\covmat{\Eb}$ given in Lemma~\ref{lem:GaussRVError}, we observe that $\covmat{\Eb}=\covmat{\Xb|\Yb}$ if the matrix $T$ in Lemma~\ref{lem:GaussRVError} satisfies $\covmat{\Yb\Xb}=\covmat{\Yb} \trans{T}$. This is indeed the case: since $\Xb = \bfmu+T\Yb+\Eb$, and since $\Yb$ and $\Eb$ are uncorrelated, we have that
\begin{IEEEeqnarray}{lCl}
 \covmat{\Yb\Xb} &=& \expec{\Yb\trans{\Xb}} - \expec{\Yb}\expec{\trans{\Xb}} \nonumber\\
&=&  \expec{\Yb\trans{(\bfmu+T\Yb+\Eb)}} - \expec{\Yb}\expec{\trans{(\bfmu+T\Yb+\Eb)}}\nonumber\\
 &=&  \expec{\Yb}\trans{\bfmu} + \expec{\Yb\trans{\Yb}}\trans{T}-\expec{\Yb}\trans{\bfmu}-\expec{\Yb}\expec{\trans{\Yb}}\trans{T}\nonumber\\ 
&=&  \covmat{\Yb}\trans{T}.
\end{IEEEeqnarray}
This proves Theorem~\ref{thm:dim:condGauss}.

\section{Proof of Lemma~\ref{lem:rate:lipschitz}}\label{proofs:dim:lipschitz}
To prove Lemma~\ref{lem:rate:lipschitz}, we shall need the following auxiliary result.
\begin{lem}\label{lem:Lipschitz}
 Let $X_1^k$ be a collection of real-valued RVs, and let $f\colon \reals^k\to\reals^\ell$ be Lipschitz continuous with Lipschitz constant $\lipschitz$. Then,
 \begin{equation}
  \ent{\qRVm{f(X_1^k)}|\qRVm{X_1^k}} \le \ell \log \lceil \lipschitz \sqrt{k} +1\rceil.
 \end{equation}
\end{lem}
\begin{IEEEproof}
 Note that if $\qRVm{X_1^k}=z_1^k/m$ for some $z_1^k\in\integers^k$, then $X_1^k\in\dom{C}(z_1^k/m,1/m)$, a cube with diameter $\sqrt{k}/m$. The image of this cube under the Lipschitz function $f$ has a diameter not greater than $\lipschitz\sqrt{k}/m$. Computing $\qRVm{f(X_1^k)}$ induces a partition of $\reals^\ell$ into $\ell$-dimensional cubes. Of this partition, at most $\lceil \lipschitz \sqrt{k} + 1\rceil^\ell$ elements have a nonempty intersection with the image of $\dom{C}(z_1^k/m,1/m)$ under $f$. Therefore,
 \begin{equation}
  \ent{\qRVm{f(X_1^k)}|\qRVm{X_1^k}=z_1^k/m} \le \ell \log \lceil \lipschitz \sqrt{k} +1\rceil
 \end{equation}
for every $z_1^k\in\integers^k$, so Lemma~\ref{lem:Lipschitz} follows by averaging over $\qRVm{X_1^k}$.
\end{IEEEproof}

We next prove Lemma~\ref{lem:rate:lipschitz}. Let $\Yb_t=f_t(\Xb_t)$. To prove the right-most relation in \eqref{eq:rate:lipschitz}, we use that for every $k$ and $m$
\begin{IEEEeqnarray}{lCl}
\ent{\qRVm{\Xb_1^k}} & \leq & \ent{\qRVm{\Xb_1^k},\qRVm{\Yb_1^k}} \nonumber\\
 & = & \ent{\qRVm{\Xb_1^k}} + \ent{\qRVm{\Yb_1^k}|\qRVm{\Xb_1^k}}. \label{eq:62}
\end{IEEEeqnarray}
The second summand can be further upper-bounded by
\begin{equation}
\label{eq:63}
 \ent{\qRVm{\Yb_1^k}|\qRVm{\Xb_1^k}} \le \sum_{t=1}^k \ent{\qRVm{f_t(\Xb_t)}|\qRVm{\Xb_t}}.
\end{equation}
Since every function $f_t$ is Lipschitz with a Lipschitz constant at most $\lipschitz\triangleq\sup_{t\in\integers} \lipschitz_t $, we can use Lemma~\ref{lem:Lipschitz} to bound the RHS of \eqref{eq:63} by $kM \log \lceil \lipschitz \sqrt{L} +1\rceil$. Since this term is independent of $m$, the contribution of the second summand on the RHS of \eqref{eq:62} vanishes as $m\to\infty$. We thus obtain $\infodimlu{\{\Xb_t,\Yb_t\}}=\infodimlu{\{\Xb_t\}}$ by dividing both sides of \eqref{eq:62} by $k \log m$ and letting $k$ and $m$ tend to infinity.

To prove the left-most relation in \eqref{eq:rate:lipschitz}, we use that for every $k$ and $m$
\begin{equation}
\label{eq:64}
\ent{\qRVm{\Yb_1^k}}\le\ent{\qRVm{\Yb_1^k},\qRVm{\Xb_1^k}}.
\end{equation}
The claim follows then by dividing both sides of \eqref{eq:64} by $k \log m$ and letting $k$ and $m$ tend to infinity.

\section{Proof of Lemma~\ref{lem:rate:mixture}}
\label{app:dim:mixture}
For every $m$ and $k$, we have
 \begin{IEEEeqnarray}{lCl}
  \ent{\qRVm{\Xb_1^k}} & \le & \ent{\qRVm{\Xb_1^k},\qRVm{\Zb_1^k}}\nonumber\\
   & \le & \ent{\qRVm{\Xb_1^k}}+\ent{\qRVm{\Zb_1^k}}. \label{eq:the_other_28}
 \end{IEEEeqnarray}
 Dividing by $k\log m$ and letting first $k$ and then $m$ tend to infinity yields \eqref{eq:rate:mixture:eq1}.

To prove \eqref{eq:rate:mixture:eq3}, we note that Lemma~\ref{lem:rate:lipschitz} and \eqref{eq:rate:mixture:eq1} yield $\infodimlu{\{\Xb_t+\Zb_t\}}\leq \infodimlu{\{\Xb_t\}}$. For the reverse inequality, we use \cite[Lemma~30]{Wu_PhD} and the fact that conditioning reduces entropy to obtain
\begin{IEEEeqnarray}{lCl}
 \IEEEeqnarraymulticol{3}{l}{\ent{\qRVm{\Xb_1^k+\Zb_1^k}}} \nonumber\\
 \quad & \ge &  \ent{\qRVm{\Xb_1^k+\Zb_1^k}|\qRVm{\Zb_1^k}}\nonumber\\
 & \geq & \ent{\qRVm{\Xb_1^k}+\qRVm{\Zb_1^k}|\qRVm{\Zb_1^k}} - kL\log(2) \nonumber\\
 & \geq & \ent{\qRVm{\Xb_1^k}} - \ent{\qRVm{\Zb_1^k}}- kL\log(2). \label{eq:68}
\end{IEEEeqnarray}
Dividing both sides of \eqref{eq:68} by $k\log m$, and letting first $k$ and then $m$ tend to infinity, yields $\infodimlu{\{\Xb_t+\Zb_t\}}\geq \infodimlu{\{\Xb_t\}}$ and proves \eqref{eq:rate:mixture:eq3}.

Finally, if $Z$ is discrete and $\ent{\qRV{\Xb_1}{1}}=\infty$, then $\ent{\qRVm{\Xb_1^k}|Z}=\infty$, since
\begin{equation}
 \ent{\qRVm{\Xb_1^k}|Z} \geq \ent{\qRVm{\Xb_1^k}} - \ent{Z}
\end{equation}
where the second entropy is finite by assumption and the first entropy satisfies $\ent{\qRVm{\Xb_1}}\geq \ent{\qRV{\Xb_1}{1}}=\infty$. Conversely, if $Z$ is discrete and $\ent{\qRV{\Xb_1}{1}}<\infty$, then
\begin{IEEEeqnarray}{lCl}
  \ent{\qRVm{\Xb_1^k}|Z} & \le & \ent{\qRVm{\Xb_1^k}} \nonumber\\
  & = & \ent{\qRVm{\Xb_1^k}|Z} + \mutinf{\qRVm{\Xb_1^k};Z}. \IEEEeqnarraynumspace
 \end{IEEEeqnarray}
Dividing all terms by $k$ and letting $k$ tend to infinity thus yields
\begin{IEEEeqnarray}{lCl}
 \IEEEeqnarraymulticol{3}{l}{\limk \frac{\ent{\qRVm{\Xb_1^k}|Z}}{k}} \nonumber\\
 \quad & \le & \entrate{\{\qRVm{\Xb_t}\}} \nonumber\\
 & \le & \limk \frac{\ent{\qRVm{\Xb_1^k}|Z}}{k} + \varlimsup_{k\to\infty} \frac{1}{k}\mutinf{\qRVm{\Xb_1^k};Z}. \label{eq:rate:mixture:eq}
\end{IEEEeqnarray}
Since $\mutinf{\qRVm{\Xb_1^k};Z}\leq\ent{Z}<\infty$, the second term on the RHS of~\eqref{eq:rate:mixture:eq} tends to zero as $k$ tends to infinity. Thus, dividing~\eqref{eq:rate:mixture:eq} by $\log m$, and letting $m$ tend to infinity, yields \eqref{eq:rate:mixture:eq2}.

\section{Proof of Theorem~\ref{thm:rate:rddim}}\label{proof:rddim}
\newcommand{\sqeucnorm}[1]{\|#1\|^2}
The proof of Theorem~\ref{thm:rate:rddim} is essentially identical to the proof of \cite[Lemma~3.2]{Kawabata_RDDim}. For the sake of completeness, we reproduce the full proof here. Indeed, choosing in \eqref{eq:R2(D)}
\begin{equation}
\label{eq:Xhat}
\hat{\Xb}_1^k = \qRV{\Xb_1^k}{m}, \quad m=\sqrt{\frac{L}{D}}
\end{equation}
yields
\begin{equation}
R(\Xb_1^k,kD) \leq \ent{\qRV{\Xb_1^k}{m}}
\end{equation}
since for the choice \eqref{eq:Xhat} we have $\|\Xb_1^k-\hat{\Xb}_1^k\|_2^2\leq \frac{kL}{m^2}=kD$, hence it satisfies \eqref{eq:D}. Consequently, dividing by $-k\log D$, and taking limits as $k\to\infty$ and $D\downarrow 0$, we obtain
\begin{align}
2 \lim_{D\downarrow 0}\limk \frac{R(\Xb_1^k,kD)}{-k\log D} 
& \leq 2\lim_{D\downarrow 0}\limk \frac{\ent{\qRV{\Xb_1^k}{m}}}{-k\log D} \nonumber\\
& =  \lim_{m\to\infty} \lim_{k\to\infty} \frac{\ent{\qRV{\Xb_1^k}{m}}}{k\log m} \label{eq:UB_limsup}
\end{align}
if the limits exist. If the limits do not exist, then we obtain the same upper bound for the limits replaced by limits superior and limits inferior.\footnote{Since $m^2=L/D$, taking the limit as $D\downarrow 0$ is tantamount to taking the limit as $m\to\infty$.}

We next derive a lower bound on the rate-distortion dimension. To simplify notation, we treat the collection $\Xb_1^k$ of $k$ $L$-variate random vectors as a collection of $k'=kL$ RVs. To show that the upper bound~\eqref{eq:UB_limsup} holds with equality, we use the following lower bound on $R(\Xb_1^k,D)$ given in \cite{Csiszar}, \cite[eq.~(A.1)]{Kawabata_RDDim}:
\begin{equation}
\label{eq:dembo}
R(X_1^{k'},D) \geq \sup_{s\leq 0, \lambda_s} \left\{s D + \expec{\log \lambda_s\left(X_1^{k'}\right)}\right\}
\end{equation}
where $\lambda_s\colon \reals^{k'}\to [0,\infty)$ is an arbitrary nonnegative measurable function satisfying
\begin{equation}
\label{eq:lambdas}
\sup_{y_1^{k'}\in\reals^{k'}} \expec{\lambda_s\left(X_1^{k'}\right) \e{s \sqeucnorm{y_1^{k'}-X_1^{k'}}}} \leq 1.
\end{equation}
Following the proof of \cite[Lemma~3.2]{Kawabata_RDDim}, we apply \eqref{eq:dembo} with
\begin{align}
s & =  - m^2 \label{eq:s_Tobi}\\
\lambda_s(x_1^{k'}) & =  \frac{1}{N^{k'}} \sum_{i_1^{k'}\in\integers^{k'}} \frac{\I{\qRV{x_1^{k'}}{m}=\frac{i_1^{k'}}{m}}}{\Prob(\qRV{X_1^{k'}}{m}=\frac{i_1^{k'}}{m})} \label{eq:lambdas_Tobi}\\
N & =  1 + 2\sum_{i=0}^{\infty} \e{-i^2}. \label{eq:N_Tobi}
\end{align}
We first show that this choice of $\lambda_s$ satisfies \eqref{eq:lambdas}. Indeed, 
\begin{IEEEeqnarray}{lCl}
\IEEEeqnarraymulticol{3}{l}{\sup_{y_1^{k'}\in\reals^{k'}} \expec{\lambda_s\left(X_1^{k'}\right) \e{s \sqeucnorm{y_1^{k'}-X_1^{k'}}}}} \nonumber\\
& \leq & \sup_{y_1^{k'}\in\reals^{k'}} \frac{1}{N^{k'}} \sum_{i_1^{k'}\in\integers^{k'}} \sup_{x_1^{k'}:\qRV{x_1^{k'}}{m}=\frac{i_1^{k'}}{m}} \e{-m^2 \sqeucnorm{y_1^{k'}-x_1^{k'}}} \nonumber\\
& = & \frac{1}{N^{k'}} \sup_{j_1^{k'}\in\integers^{k'}}\sup_{\tilde{y}_1^{k'}\in [0,1)^{k'}} \sum_{i_1^{k'}\in\integers^{k'}} \prod_{\ell=1}^{k'} \sup_{0\leq \tilde{x}_{\ell}<1} \e{-(\tilde{y}_{\ell}+j_{\ell}-\tilde{x}_{\ell}-i_{\ell})^2} \nonumber\\
& = & \frac{1}{N^{k'}}\prod_{\ell=1}^{k'} \sup_{j_{\ell}\in\integers}\sup_{0\leq\tilde{y}_{\ell}<1}\sum_{i_{\ell}\in\integers} \sup_{0\leq \tilde{x}_{\ell}<1} \e{-(\tilde{y}_{\ell}+j_{\ell}-\tilde{x}_{\ell}-i_{\ell})^2}
\end{IEEEeqnarray}
where the second step follows by substituting $\tilde{x}_{\ell}=m x_{\ell} - i_{\ell}$ and $\tilde{y}_{\ell}=m y_{\ell}-j_{\ell}$. Since the sum over $i_{\ell}$ does not depend on $j_{\ell}$, it follows that
 \begin{multline}
 \sup_{j_{\ell}\in\integers}\sup_{0\leq\tilde{y}_{\ell}<1}\sum_{i_{\ell}\in\integers} \sup_{0\leq \tilde{x}_{\ell}<1} \e{-(\tilde{y}_{\ell}+j_{\ell}-\tilde{x}_{\ell}-i_{\ell})^2}
 \\
 = \sup_{0\leq\tilde{y}_{\ell}<1}\sum_{i_{\ell}\in\integers} \sup_{0\leq \tilde{x}_{\ell}<1} \e{-(\tilde{y}_{\ell}-\tilde{x}_{\ell}-i_{\ell})^2}
 \end{multline}
which can be upper-bounded as
\begin{equation}
\sup_{0\leq\tilde{y}_{\ell}<1}\sum_{i_{\ell}\in\integers} \sup_{0\leq \tilde{x}_{\ell}<1} \e{-(\tilde{y}_{\ell}-\tilde{x}_{\ell}-i_{\ell})^2} \leq 1 + 2 \sum_{i=0}^{\infty} \e{- i^2}.
\end{equation}
Hence,
\begin{multline}
\sup_{y_1^{k'}\in\reals^{k'}} \expec{\lambda_s\left(X_1^{k'}\right) \e{s \sqeucnorm{y_1^{k'}-X_1^{k'}}}} \\\leq \frac{1}{N^{k'}} \left(1 + 2 \sum_{i=0}^{\infty} \e{- i^2}\right)^{k'}
\end{multline}
which, by \eqref{eq:N_Tobi}, is equal to $1$. It follows that $s$ and $\lambda_s$, as chosen in \eqref{eq:s_Tobi} and \eqref{eq:lambdas_Tobi}, satisfy \eqref{eq:lambdas}.

We next evaluate \eqref{eq:dembo} for this choice of $s$ and $\lambda_s$ and for distortion $kD$. This yields
\begin{IEEEeqnarray}{lCl}
R(X_1^{k'},kD)
& \geq & -k m^2 D -k' \log N \nonumber\\
& & {} + \expec{\log\left(\sum_{i_1^{k'}\in\integers^{k'}}\frac{\I{\qRV{X_1^{k'}}{m}=\frac{i_1^{k'}}{m}}}{\Prob\left(\qRV{X_1^{k'}}{m}=\frac{i_1^{k'}}{m}\right)}\right)} \nonumber\\
& = & \ent{\qRV{X_1^{k'}}{m}} - k(m^2 D + L\log N).
\end{IEEEeqnarray}
For $m^2 = L/D$, this becomes
\begin{equation}
\label{eq:rddim_164}
R(X_1^{k'},kD) \geq \ent{\qRV{X_1^{k'}}{m}} - k'(1 + \log N).
\end{equation}

We next replace again the collection $X_1^{k'}$ of RVs by the equivalent collection $\Xb_1^k$ of random vectors. Dividing both sides of \eqref{eq:rddim_164} by $-k\log D$, and taking the limits as $k\to\infty$ and $D\downarrow 0$, yields
\begin{IEEEeqnarray}{lCl}
\IEEEeqnarraymulticol{3}{l}{2 \lim_{D\downarrow 0} \limk \frac{R(\Xb_1^{k},kD)}{-k\log D}}\nonumber\\
\quad & \geq & 2 \lim_{D\downarrow 0} \limk \frac{\ent{\qRV{\Xb_1^{k}}{m}} - kL(1 + \log N)}{-k\log D} \nonumber\\
& = & \lim_{m\to \infty} \limk \frac{\ent{\qRV{\Xb_1^{k}}{m}}}{k\log m} \label{eq:LB_limsup}
\end{IEEEeqnarray}
if the limits over $k$ and $D$ exist. If the limits do not exist, then we obtain the same lower bound for the limits replaced by limits superior and limits inferior. Combining \eqref{eq:LB_limsup} with \eqref{eq:UB_limsup} proves Theorem~\ref{thm:rate:rddim}.

\section{Proof of Theorem~\ref{thm:rate:Gauss}}\label{proofs:rate:Gauss}
The proof consists of two parts. In the first part, we show that of all processes $\{\Xb_t\}$ with a given SDF $\sdf{\Xb}$, the Gaussian process has the largest information dimension rate (Section~\ref{sub:GaussMax}). In the second part, we demonstrate that the information dimension rate of Gaussian processes is given by the average rank of the derivative of the SDF (Section~\ref{sub:InfoDimGauss}).

\subsection{Gaussian Processes Maximize the Information Dimension}\label{sub:GaussMax}
By Theorem~\ref{thm:rate:rddim}, the upper information dimension rate is given by
\begin{equation}
\label{eq:bjbjbjb}
\infodimu{\{\Xb_t\}} = 2\varlimsup_{D\downarrow 0} \lim_{k\to\infty} \frac{R(\Xb_1^k,kD)}{-k \log D}.
\end{equation}
The claim that the information dimension is maximized by a Gaussian process then follows by the well-known fact that of all random vectors $\Xb_1^k$ with a given covariance matrix $\covmat{\Xb_1^k}$, the Gaussian random vector has the largest rate-distortion function $R(\Xb_1^k,kD)$.

To prove this claim for multivariate sources, we shall write the collection of $L$-variate vectors $\Xb_1^k$ as a collection of $k'$ RVs $X_1^{k'}$, where $k'=kL$. Since the information dimension rate is translation invariant (Lemma~\ref{lem:rate:lipschitz}), we can assume without loss of optimality that the RVs $X_1^{k'}$ have zero mean. Furthermore, by the eigenvalue decomposition, there exists an orthogonal matrix $W$ such that the random variables $Y_1^{k'}$ given by $\trans{(Y_1,\dots,Y_{k'})}= \trans{W}\trans{(X_1,\ldots,X_{k'})}$ are uncorrelated and their variances are the eigenvalues of $\covmat{X_1^{k'}}$, which we shall denote by $\lambda_1,\ldots,\lambda_{k'}$. Since mutual information is invariant under bijections, and the Euclidean norm is invariant under multiplications by orthogonal matrices, it follows that $R(Y_1^{k'},kD)=R(X_1^{k'},kD)$.

For the case where $Y_1^{k'}$ are independent, zero-mean, Gaussian random variables with variances $\lambda_1,\ldots,\lambda_{k'}$, the rate-distortion function is given by \cite[Th.~13.3.3]{Cover_Information1}
\begin{equation}
\label{eq:RD_Gauss}
R(Y_1^{k'},kD) = \sum_{t=1}^{k'} \frac{1}{2}\log\frac{\lambda_t}{D_t}
\end{equation}
where
\begin{equation}
\label{eq:xi}
D_t = \begin{cases} \xi, & \textnormal{if $\xi<\lambda_t$} \\ \lambda_t, & \textnormal{if $\xi\geq \lambda_t$} \end{cases}
\end{equation}
and $\xi$ is chosen so that $D_1+\ldots+D_{k'}=kD$.

The RHS of \eqref{eq:RD_Gauss} can also be achieved for non-Gaussian RVs by choosing the following (possibly suboptimal) distribution of the reconstruction values $\hat{Y}_1^{k'}$:
\begin{equation}
\label{eq:This_Dist}
\hat{Y}_t = \begin{cases} \left(Y_t+Z_t\right)\frac{\lambda_t-D_t}{\lambda_t}, & \textnormal{if $\xi<\lambda_t$}\\ 0, & \textnormal{if $\xi\geq \lambda_t$}\end{cases}
\end{equation}
where $Z_1^{k'}$ are independent, zero-mean, Gaussian RVs with variances $\frac{D_t\lambda_t}{\lambda_t-D_t}$, and $\xi$ is as in \eqref{eq:xi}. Indeed, it is easy to check that \eqref{eq:This_Dist} satisfies the distortion constraint \eqref{eq:D}. Furthermore, by using that conditioning reduces entropy and that Gaussian RVs maximize differential entropy, it can be shown that
\begin{equation}
I(\hat{Y}_1^{k'};Y_1^{k'}) \leq \sum_{t=1}^{k'}\frac{1}{2} \log\frac{\lambda_t}{D_t}. \label{eq:RD_nonGauss}
\end{equation}
Comparing \eqref{eq:RD_nonGauss} with \eqref{eq:RD_Gauss}, we conclude that, for uncorrelated RVs $Y_1^{k'}$,
\begin{equation}
R(Y_1^{k'},kD) \leq R\bigl((Y_1^{k'})_G,kD\bigr)
\end{equation}
where $(Y_1^{k'})_G$ are jointly Gaussian with the same covariance matrix as $Y_1^{k'}$. Since $R(Y_1^{k'},kD)=R(X_1^{k'},kD)$ and $R\bigl((Y_1^{k'})_G,kD\bigr)=R\bigl((X_1^{k'})_G,kD\bigr)$, the same is also true for general RVs. Together with \eqref{eq:bjbjbjb}, this proves that of all processes $\{\Xb_t\}$ with a given SDF $\sdf{\Xb}$, the Gaussian process has the largest information dimension rate.

\subsection{The Information Dimension of Gaussian Processes}\label{sub:InfoDimGauss}
We now assume that $\{\Xb_t\}$ is Gaussian. For every $i$, we define \begin{equation}
N_{i,t}\triangleq X_{i,t}-\qRVm{X_{i,t}}.
\end{equation}
Furthermore, let $U_{i,t}$ be i.i.d. (over all $i$ and $t$) and uniformly distributed on $[0,1/m)$, and let $W_{i,t} \triangleq \qRVm{X_{i,t}}+U_{i,t}$. We define $\{\qRVm{\Xb_{t}}\}$, $\{\Nb_t\}$, and $\{\Ub_t\}$ as the corresponding multivariate processes. Since $\{U_{i,t}\}$ is independent of $\{\qRVm{X_{j,t}}\}$ for every $i,j$, the (matrix-valued) SDFs of $\{\Wb_t\}$, $\{\qRVm{\Xb_{t}}\}$, and $\{\Ub_t\}$ satisfy
\begin{equation}
\label{eq:T_sdftW}
\sdft{\Wb}=\sdft{\qRVm{\Xb}}+\sdft{\Ub}.
\end{equation}
Moreover, the (matrix-valued) PSD of $\{\Ub_t\}$ exists and equals
\begin{equation}
\label{eq:T_sdftU}
\psdt{\Ub}=\frac{1}{12m^2} I_L.
\end{equation}

Since the information dimension rate is translation invariant (Lemma~\ref{lem:rate:lipschitz}), and since the SDF $\sdf{\Xb}$ does not depend on the mean vector $\bfmu$, we can assume without loss of generality that $\{\Xb_t\}$ has zero mean. We further show in Lemma~\ref{lem:thm:unitvariance} in Appendix~\ref{app:aux} that we can assume without loss of generality that every component process of $\{\Xb_t\}$ has unit variance. By~\eqref{eq:SDMSumEqual} in Lemma~\ref{lem:thm:bussgang}, it thus follows that
\begin{equation}
\label{eq:T_sdftRVm}
\sdft{\qRVm{\Xb}} = (2a_1-1)\sdft{\Xb} + \sdft{\Nb}.
\end{equation}

We continue by writing the entropy of $\qRVm{\Xb_1^k}$ in terms of a differential entropy, i.e.,
 \begin{equation}
  \ent{\qRVm{\Xb_1^k}} = \diffent{\Wb_1^k}+kL\log m.
 \end{equation}
Denoting by $(\Wb_1^k)_G$ a Gaussian vector with the same mean and covariance matrix as $\Wb_1^k$, and denoting by $\pdf{\Wb_1^k}$ and $\pdfg{\Wb_1^k}$ the PDFs of $\Wb_1^k$ and $(\Wb_1^k)_G$, respectively, this can be expressed as
 \begin{equation}
 \label{eq:derate_quantent}
  H(\qRV{\Xb_1^k}{m}) = \diffent{(\Wb_1^k)_G} - \kld{\pdf{\Wb_1^k}}{\pdfg{\Wb_1^k}} + kL\log m.
 \end{equation}
Dividing by $k\log m$, and letting first $k$ and then $m$ tend to infinity, yields the information dimension rate $d(\{\Xb_t\})$. Lemma~\ref{lem:thm:approxbound} in Appendix~\ref{app:aux} shows that
\begin{equation}
\kld{\pdf{\Wb_1^k}}{\pdfg{\Wb_1^k}} \leq k\Xi
\end{equation}
for some constant $\Xi$ that is independent of $(k,m)$. Moreover, the differential entropy rate of the stationary, $L$-variate, Gaussian process $(\{\Wb_t\})_G$ is given by~\cite[Th.~7.10]{Wiener_StochasticProcesses}
  \begin{multline}\label{eq:derate_sdm}
   \limk \frac{\diffent{(\Wb_1^k)_G}}{k}  \\ = \frac{L}{2}\log (2\pi e) + \frac{1}{2} \int_{-1/2}^{1/2} \log\det\dsdft{\Wb} \diff\theta.
  \end{multline}
It thus follows that the information dimension rate of $\{\Xb_t\}$ equals
\begin{equation}\label{eq:proof:Lplus}
 d(\{\Xb_t\}) = L + \limm \frac{1}{2\log m} \int_{-1/2}^{1/2} \log\det\dsdft{\Wb} \diff\theta.
\end{equation}
It remains to show that the RHS of~\eqref{eq:proof:Lplus} is equal to the RHS of~\eqref{eq:thm:mainline}. To do so, we first show that the integral on the RHS of~\eqref{eq:proof:Lplus} can be restricted to a subset $\dom{F}_\Upsilon^\mathsf{c}\subseteq[-1/2,1/2]$ on which the entries of $\dsdft{\Nb}$ are bounded from above by $\Upsilon/m^2$ for some $\Upsilon>0$. We then show that, on this set, $\det\dsdft{\Wb}$ can be bounded from above and from below by products of affine transforms of the \emph{eigenvalues of $\dsdft{\Xb}$}. These bounds are asymptotically tight, i.e., they are equal in the limit as $m$ tends to infinity. We complete the proof by showing that the order of limit and integration can be exchanged.

\subsubsection{Restriction on $\dom{F}_\Upsilon^\mathsf{c}\subseteq[-1/2,1/2]$}
Choose $\Upsilon>0$ and let
\begin{IEEEeqnarray}{lCl}
\mathcal{F}^{(i)}_{\Upsilon} & \triangleq & \left\{\theta\colon \dsdft{N_{i}}> \Upsilon/m^2\right\} \\
 \dom{F}_\Upsilon & \triangleq & \{\theta\colon \max_{i=1,\dots,L} \dsdft{N_{i}}> \Upsilon/m^2\}.
\end{IEEEeqnarray}
By \eqref{eq:boundNoise} in Lemma~\ref{lem:thm:bussgang}, we have for every $i$
 \begin{equation}
 \label{eq:sdfX-Z}
  \lebesgue\left(\mathcal{F}^{(i)}_{\Upsilon}\right) \le \frac{1}{\Upsilon}.
 \end{equation}
  Since the set $\dom{F}_\Upsilon$ is the union of $\mathcal{F}^{(i)}_{\Upsilon}$, $i=1,\ldots,L$, it then follows by the union bound that
\begin{equation}
\label{eq:domFleU}
 \lebesgue(\dom{F}_\Upsilon)\le\frac{L}{\Upsilon}.
 \end{equation}
To prove \eqref{eq:sdfX-Z}, we note that, by the Lebesgue decomposition theorem \cite[Th.~2.2.6]{AsDo00} and the fact that the Radon-Nikodym derivative $\diff \sdft{N_i}/\diff \lambda$ coincides with $\dsdft{N_{i}}$ almost everywhere \cite[Sec.~2.3]{AsDo00},
\begin{IEEEeqnarray}{lCl}
\int_{-1/2}^{1/2} \diff \sdft{N_{i}} & \geq & \int_{-1/2}^{1/2} \dsdft{N_{i}} \diff\theta \nonumber\\
& \geq & \int_{\mathcal{F}^{(i)}_{\Upsilon}}\dsdft{N_{i}} \diff\theta \nonumber\\
& \geq & \frac{\Upsilon}{m^2} \lambda\left(\mathcal{F}^{(i)}_{\Upsilon}\right) \label{eq:sdfX-Z_2}
\end{IEEEeqnarray}
where the second inequality follows because $\theta\mapsto\dsdft{N_{i}}$ is nonnegative, and the third inequality follows by definition of $\mathcal{F}^{(i)}_{\Upsilon}$. By \eqref{eq:boundNoise} in Lemma~\ref{lem:thm:bussgang}, the integral on the left-hand side (LHS) of \eqref{eq:sdfX-Z_2} is upper-bounded by $1/m^2$, hence \eqref{eq:sdfX-Z} follows.
 
By \eqref{eq:T_sdftW} and \eqref{eq:T_sdftU}, we have that
\begin{equation}
\label{eq:need_this}
\dsdft{\Wb} = \dsdft{\qRVm{\Xb}} + \frac{1}{12m^2} I_L.
\end{equation}
Since derivatives of matrix-valued SDFs are positive semidefinite, it follows that
\begin{equation}
\det \dsdft{\Wb} \ge \det \psdt{\Ub} = {1}/{(12m^2)^L}.
\end{equation}
Hence,
 \begin{IEEEeqnarray}{lCl}
\varliminf_{m\to\infty} \frac{\int_{\dom{F}_\Upsilon} \log \det \dsdft{\Wb} \diff\theta}{\log m} & \ge & - \varlimsup_{m\to\infty} \lebesgue(\dom{F}_\Upsilon)\frac{ L \log(12m^2)}{\log m} \nonumber\\
& \ge & -\frac{2L^2}{\Upsilon}
 \end{IEEEeqnarray}
where the last step follows from \eqref{eq:domFleU}. Applying Hadamard's and Jensen's inequality, we further get
 \begin{IEEEeqnarray}{lCl}
\IEEEeqnarraymulticol{3}{l}{\int_{\dom{F}_\Upsilon} \log \det \dsdft{\Wb} \diff\theta} \nonumber\\
\quad  &\le & \sum_{i=1}^L  \int_{\dom{F}_\Upsilon} \log \dsdft{W_{i}} \diff\theta\nonumber\\
  & \le & \sum_{i=1}^L\lebesgue(\dom{F}_\Upsilon) \log\left(\frac{\int_{\dom{F}_\Upsilon} \dsdft{W_{i}} \diff\theta}{\lebesgue(\dom{F}_\Upsilon)}\right)\nonumber\\
  & \le & L \lebesgue(\dom{F}_\Upsilon) \log\left(\frac{(2a_1-1)+\frac{1}{12m^2}+\frac{1}{m^2}}{\lebesgue(\dom{F}_\Upsilon)}\right) \label{eq:207}
 \end{IEEEeqnarray}
 where the last step follows from \eqref{eq:boundNoise}, \eqref{eq:T_sdftW}, \eqref{eq:T_sdftRVm}, and the assumption that every component process of $\{\Xb_t\}$ has zero mean and unit variance. Since, by~\eqref{eq:boundBussgang} in Lemma~\ref{lem:thm:bussgang}, $a_1 \to 1$ as $m\to \infty$, \eqref{eq:207} yields
 \begin{IEEEeqnarray}{lCl}
  \varlimsup_{m\to\infty} \int_{\dom{F}_\Upsilon} \log \det \dsdft{\Wb} \diff\theta & \le & - \varliminf_{m\to\infty} L \lebesgue(\dom{F}_\Upsilon) \log\left(\lebesgue(\dom{F}_\Upsilon)\right) \nonumber\\
  & \leq & \frac{L}{e}.
\end{IEEEeqnarray}
Consequently,
   \begin{multline}
   -\frac{2L^2}{\Upsilon} 
 \le \varliminf_{m\to\infty} \frac{\int_{\dom{F}_\Upsilon} \log \det \dsdft{\Wb} \diff\theta}{\log m} \\
 \le \varlimsup_{m\to\infty} \frac{\int_{\dom{F}_\Upsilon} \log \det \dsdft{\Wb} \diff\theta}{\log m} \label{eq:throw_away}
 \le 0
  \end{multline}
 for every $\Upsilon$. It follows that this integral does not contribute to the information dimension rate if we let $\Upsilon$ tend to infinity. In view of \eqref{eq:proof:Lplus}, we thus obtain the information dimension rate $d(\{\Xb_t\})$ by evaluating
 \begin{equation}\label{eq:integralToEvaluate}
  L+\frac{1}{2\log m} \int_{\dom{F}^\mathsf{c}_\Upsilon} \log \det \dsdft{\Wb} \diff\theta
 \end{equation}
 in the limit as first $m$ and then $\Upsilon$ tends to infinity.

\subsubsection{Bounding $\det \dsdft{\Wb}$ by the Eigenvalues of $\dsdft{\Xb}$}
Lemma~\ref{lem:thm:bussgang} and \eqref{eq:need_this} yield
 \begin{equation}
  \dsdft{\Wb} = (2a_1-1)\dsdft{\Xb} + \dsdft{\Nb} + \frac{1}{12m^2} I_L.
 \end{equation}
Let $\chi_i(\theta)$, $i=1,\dots,L$, denote the eigenvalues of $\dsdft{\Xb}$. Since $\dsdft{\Nb}$ is positive semidefinite, we obtain
\begin{IEEEeqnarray}{lCl}
 \det \dsdft{\Wb} & \ge & \det\left((2a_1-1)\dsdft{\Xb}+ \frac{1}{12m^2} I_L\right) \nonumber\\
& = & \prod_{i=1}^L\left((2a_1-1)\chi_i(\theta) + \frac{1}{12m^2} \right). 
\label{eq:detLower}
\end{IEEEeqnarray}

We next derive an upper bound on $ \det \dsdft{\Wb}$. Let $\Vert \dsdft{\Nb}\Vert_1 \triangleq \sum_{i,j=1}^n |\dsdft{N_iN_j}|$ denote the \emph{$\ell_1$-matrix norm} of $\dsdft{\Nb}$. Since $\dsdft{\Nb}$ is positive semidefinite, the element with the maximum modulus is on the main diagonal; cf.~\cite[Problem~7.1.P1]{Horn_Matrix}. Furthermore, by assumption, on $\dom{F}_\Upsilon^\mathsf{c}$ the diagonal elements of $\sdft{\Nb}$ are bounded from above by $\frac{\Upsilon}{m^2}$. We hence obtain that
\begin{equation}
\label{bla}
\Vert \dsdft{\Nb} \Vert_1 \le L^2\frac{\Upsilon}{m^2}.
\end{equation}
It is known that all matrix norms bound the largest eigenvalue of the matrix from above~\cite[Th.~5.6.9]{Horn_Matrix}.\footnote{This bound holds without a multiplicative constant, since the spectral radius of a matrix is the infimum of all matrix norms~\cite[Lemma~5.6.10]{Horn_Matrix}.} Thus, the upper bound \eqref{bla} is also an upper bound on the largest eigenvalue of $\dsdft{\Nb}$. Let $\omega_i(\theta)$, $i=1\dots,L$, denote the eigenvalues of $\dsdft{\Wb}$. Then, we have for $m^2\geq 8/\pi$ (such that $2a_1-1\ge 0$) \cite[Cor.~4.3.15]{Horn_Matrix}
\begin{IEEEeqnarray}{lCl}
 \det \dsdft{\Wb} &= & \prod_{i=1}^L \omega_i(\theta)\notag \\
&\le & \prod_{i=1}^L \left( (2a_1-1)\chi_i(\theta) + \frac{L^2\Upsilon}{m^2} + \frac{1}{12m^2}\right). \IEEEeqnarraynumspace\label{eq:detUpper}
\end{IEEEeqnarray}
Combining~\eqref{eq:detLower} and~\eqref{eq:detUpper} with~\eqref{eq:integralToEvaluate}, we obtain
\begin{IEEEeqnarray}{lCl}
\IEEEeqnarraymulticol{3}{l}{\limm \sum_{i=1}^L\frac{\int_{\dom{F}^\mathsf{c}_\Upsilon} \log\left((2a_1-1)\chi_i(\theta) + \frac{1}{12m^2} \right) \diff\theta}{\log m}} \nonumber\\
&\le & \varliminf_{m\to\infty} \frac{\int_{\dom{F}^\mathsf{c}_\Upsilon} \log \det \dsdft{\Wb} \diff\theta}{\log m} \nonumber\\
 &\le & \varlimsup_{m\to\infty} \frac{\int_{\dom{F}^\mathsf{c}_\Upsilon} \log \det \dsdft{\Wb} \diff\theta}{\log m} \nonumber\\
 &\le & \limm \sum_{i=1}^L\frac{\int_{\dom{F}^\mathsf{c}_\Upsilon} \log\left((2a_1-1)\chi_i(\theta) + \frac{\frac{1}{12}+L^2\Upsilon}{m^2} \right) \diff\theta}{\log m}.\IEEEeqnarraynumspace\label{eq:T_lmost_rmost}
\end{IEEEeqnarray}
To compute the limit of \eqref{eq:integralToEvaluate} as $m\to\infty$, we thus need to evaluate 
 \begin{equation}\label{eq:integralReformulated}
L-\sum_{i=1}^L \limm \int_{\dom{F}^\mathsf{c}_\Upsilon} \frac{\log \left((2a_1-1)\chi_i(\theta)  + \frac{\mathsf{K}}{m^2} \right)}{\log(1/m^2)} \diff\theta 
 \end{equation}
where $\mathsf{K}$ is either $1/12$ (left-most inequality in \eqref{eq:T_lmost_rmost}) or $1/12+L^2\Upsilon$ (right-most inequality in \eqref{eq:T_lmost_rmost}).

\subsubsection{Exchanging Limit and Integration}
To evaluate \eqref{eq:integralReformulated}, we continue along the lines of~\cite[Sec.~VIII]{lapidoth05}. Specifically, for each $i$, we split the integral on the RHS of~\eqref{eq:integralReformulated} into three parts: 
 \begin{align}
  \dom{F}_I &\triangleq \{\theta\in\dom{F}_\Upsilon^\mathsf{c}{:}\ \chi_i(\theta)=0\}\\
\dom{F}_{II} &\triangleq \{\theta\in\dom{F}_\Upsilon^\mathsf{c}{:}\ \chi_i(\theta)\ge \mathsf{K}/(1-\varepsilon)\}\\
\dom{F}_{III} &\triangleq \{\theta\in\dom{F}_\Upsilon^\mathsf{c}{:}\ 0<\chi_i(\theta)<\mathsf{K}/(1-\varepsilon)\}
 \end{align}
where $0<\varepsilon< 1$ is arbitrary.

For the first part, we obtain
\begin{IEEEeqnarray}{lCl}
\IEEEeqnarraymulticol{3}{l}{\int_{\dom{F}_I} \frac{\log \left((2a_1-1)\chi_i(\theta)  + \frac{\mathsf{K}}{m^2} \right)}{\log(1/m^2)} \diff\theta } \nonumber\\
\quad & = & \int_{\dom{F}_I} \frac{\log \mathsf{K} + \log(1/m^2)}{\log(1/m^2)} \diff\theta \nonumber\\
& = & \lebesgue(\dom{F}_I)\left(1+\frac{\log \mathsf{K}}{\log(1/m^2)}\right)\label{eq:F1}
\end{IEEEeqnarray}
which evaluates to $\lebesgue(\dom{F}_I)$ in the limit as $m\to \infty$.

We next show that the integrals over $\dom{F}_{II}$ and $\dom{F}_{III}$ do not contribute to~\eqref{eq:integralReformulated}. To this end, it suffices to consider the integral of the function
\begin{equation}\label{eq:AmosFunction}
 \frac{\log \left((2a_1-1)\frac{\chi_i(\theta)}{\mathsf{K}}  + \frac{1}{m^2} \right)}{\log(1/m^2)} \triangleq \frac{\log \left(A_m(\theta)  + \frac{1}{m^2} \right)}{\log(1/m^2)}.
\end{equation}
In the remainder of the proof, we shall assume without loss of generality that $m^2> 8/\pi$, in which case $A_m(\theta)>0$ on $\theta\in\dom{F}_{II}\cup\dom{F}_{III}$. Clearly, whenever $A_m(\theta)>0$, the function in \eqref{eq:AmosFunction} converges to zero as $m\to \infty$. Moreover, for $A_m(\theta)\ge 1$, this function is nonpositive.

For all $\theta\in\dom{F}_{II}$ we have $A_m(\theta)\ge(2a_1-1)/(1-\varepsilon)$, hence we can find a sufficiently large $m_0$ such that, by \eqref{eq:boundBussgang} in Lemma~\ref{lem:thm:bussgang}, we have $A_m(\theta)\ge 1$, $m\geq m_0$. Since by the same result we also have $2a_1-1\le 2$, $m^2>8/\pi$, it follows that, for $m > \max\{m_0,\sqrt{8/\pi}\}$,
\begin{equation}
 \frac{\log \left( 2\frac{\chi_i(\theta)}{\mathsf{K}}  + \frac{1}{m^2} \right)}{\log(1/m^2)}
\le \frac{\log \left(A_m(\theta)  + \frac{1}{m^2} \right)}{\log(1/m^2)} \le 0. \label{eq:this}
\end{equation}
The LHS of \eqref{eq:this} is nonpositive and monotonically increases to zero as $m\to \infty$.  We can thus apply the monotone convergence theorem \cite[Th.~1.6.7, p.~49]{AsDo00} to get
\begin{align}
 0 &\ge \varlimsup_{m\to\infty} \int_{\dom{F}_{II}}  \frac{\log \left(A_m(\theta)  + \frac{1}{m^2} \right)}{\log(1/m^2)}\diff\theta  \notag \\
& \ge \varliminf_{m\to\infty} \int_{\dom{F}_{II}}  \frac{\log \left(A_m(\theta)  + \frac{1}{m^2} \right)}{\log(1/m^2)}\diff\theta  \notag \\
&\ge\limm \int_{\dom{F}_{II}}  \frac{\log \left(2\frac{\chi_i(\theta)}{\mathsf{K}}  + \frac{1}{m^2} \right)}{\log(1/m^2)}\diff\theta\notag\\
&=  \int_{\dom{F}_{II}} \limm \frac{\log \left( 2\frac{\chi_i(\theta)}{\mathsf{K}}  + \frac{1}{m^2} \right)}{\log(1/m^2)}\diff\theta \notag\\
&=0. \label{eq:F2}
\end{align}

We next turn to the case $\theta\in\dom{F}_{III}$. It was shown in~\cite[p.~443]{lapidoth05} that if $A_m(\theta)<1$, then the function in~\eqref{eq:AmosFunction} is bounded from above by 1. Furthermore, if $A_m(\theta)<1-\frac{1}{m^2}$ then it is nonnegative, and if $A_m(\theta)\geq 1-\frac{1}{m^2}$ then it is nonpositive and monotonically increasing in $m$. Restricting ourselves to the case $m^2>8/\pi$, we thus obtain for $\theta\in\dom{F}_{III}$
\begin{equation}
 \frac{\log \left(A_m(\theta)  + \frac{1}{m^2} \right)}{\log(1/m^2)} \ge 
 \begin{cases}
      \frac{\log \left(\frac{2}{1-\varepsilon}  + \frac{\pi}{8} \right)}{\log(\pi/8)}, & A_m(\theta)\ge 1-\frac{\pi}{8}\\
      0, & \text{otherwise}
 \end{cases}
\end{equation}
where we made use of the fact that $A_m(\theta)<(2a_1-1)/(1-\varepsilon)$, $\theta\in\dom{F}_{III}$ and, by \eqref{eq:boundBussgang} in Lemma~\ref{lem:thm:bussgang}, $2a_1-1\leq 2$, $m^2>8/\pi$. Hence, on $\dom{F}_{III}$ the magnitude of the function in \eqref{eq:AmosFunction} is bounded by
\begin{equation}\label{eq:AmosBound}
 \left|\frac{\log \left(A_m(\theta)  + \frac{1}{m^2} \right)}{\log(1/m^2)}\right| 
 \le \max\left\{1,\frac{\log \left(\frac{2}{1-\varepsilon}  + \frac{\pi}{8} \right)}{\log(8/\pi)}\right\}.
\end{equation}
We can thus apply the dominated convergence theorem \cite[Th.~1.6.9, p.~50]{AsDo00} to get
 \begin{multline}\label{eq:F3}
   \limm \int_{\dom{F}_{III}}  \frac{\log \left((2a_1-1)\chi_i(\theta)  + \frac{\mathsf{K}}{m^2} \right)}{\log(1/m^2)}\diff\theta\\
  =  \int_{\dom{F}_{III}} \limm \frac{\log \left((2a_1-1)\frac{\chi_i(\theta)}{\mathsf{K}}  + \frac{1}{m^2} \right)}{\log(1/m^2)}\diff\theta 
  =0.
 \end{multline}
Combining~\eqref{eq:F1},~\eqref{eq:F2}, and~\eqref{eq:F3}, we can evaluate \eqref{eq:integralReformulated} as
\begin{IEEEeqnarray}{lCl}
\IEEEeqnarraymulticol{3}{l}{L-\sum_{i=1}^L \limm \int_{\dom{F}^\mathsf{c}_\Upsilon} \frac{\log \left((2a_1-1)\chi_i(\theta)  + \frac{\mathsf{K}}{m^2} \right)}{\log(1/m^2)} \diff\theta} \nonumber\\
\quad & = & \sum_{i=1}^L \left( 1-\lebesgue(\{\theta\in\dom{F}_\Upsilon^\mathsf{c}{:}\ \chi_i(\theta)=0\})\right) \nonumber\\
& = & \sum_{i=1}^L \lebesgue(\{\theta\in\dom{F}_\Upsilon^\mathsf{c}{:}\ \chi_i(\theta)>0\}). \label{eq:after_lim}
\end{IEEEeqnarray}

\subsubsection{Wrapping Up}
To compute the limit of \eqref{eq:integralToEvaluate} as first $m$ and then $\Upsilon$ tends to infinity, it remains to let $\Upsilon\to\infty$ on the RHS of \eqref{eq:after_lim}. By the continuity of the Lebesgue measure, this yields
\begin{equation}
 \sum_{i=1}^L \lebesgue(\{\theta{:}\ \chi_i(\theta)>0\})= \int_{-1/2}^{1/2} \rank{\dsdft{\Xb}} \diff \theta.
\end{equation}

To summarize, combining \eqref{eq:proof:Lplus}, \eqref{eq:throw_away}, and \eqref{eq:after_lim}, we obtain that
\begin{IEEEeqnarray}{lCl}
 d(\{\Xb_t\}) & = & L + \lim_{\Upsilon\to\infty}\limm \frac{1}{2\log m} \int_{-1/2}^{1/2} \log\det\dsdft{\Wb} \diff\theta \nonumber\\
 & = & \int_{-1/2}^{1/2} \rank{\dsdft{\Xb}} \diff \theta.
\end{IEEEeqnarray}
This proves Theorem~\ref{thm:rate:Gauss}.

\subsection{Auxiliary Results}
\label{app:aux}
\begin{lem}\label{lem:thm:unitvariance}
 Suppose that  $\{\Xb_t\}$ is a stationary, $L$-variate, real-valued, Gaussian process with mean vector $\bfmu$ and SDF $\sdf{\Xb}$. Suppose that the component processes are ordered by their variances, i.e.,
 \begin{equation}
 \sigma_1^2\ge \sigma_2^2\ge\cdots\sigma_{L'}^2\ge\sigma_{L'+1}^2=\cdots\sigma_L^2=0.
 \end{equation}
 Then,
 \begin{equation}
 \infodim{\{\Xb_t\}}=d\left(\left\{\left(X_{1,t}/\sigma_1,\dots,X_{L',t}/\sigma_{L'}\right)\right\}\right)
 \end{equation}
 and, for almost every $\theta$,
 \begin{equation}
 \rank{\dsdft{\Xb}}=\rank{\dsdft{(X_1/\sigma_1,\dots,X_{L'}/\sigma_{L'})}}.
 \end{equation}
\end{lem}
\begin{IEEEproof}
 Normalizing component processes with positive variance to unit variance does not affect the information dimension rate, as follows from Lemma~\ref{lem:rate:lipschitz}. If $\sigma_i^2=0$, then the component process $\{X_{i,t}\}$ is almost surely constant. It follows that $\ent{\qRVm{X_{i,1}^k}}=0$ for every $m$ and every $k$, so
 \begin{equation}
 \ent{\qRVm{\Xb_1^k}} =  \ent{\qRVm{X_{1,1}^k},\dots,\qRVm{X_{L',1}^k}}.
 \end{equation}
 Dividing by $k\log m$, and letting $m$ and $k$ tend to infinity, shows that $\infodim{\{\Xb_t\}}=d(\{\frac{1}{\sigma_1}X_{1,t},\dots,\frac{1}{\sigma_{L'}}X_{L',t}\})$.

Let $\Pi$ be an $L'\times L'$ diagonal matrix with values $\sigma_i$ on the main diagonal. For component processes with zero variance, the corresponding row and column of $\dsdft{\Xb}$ is zero almost everywhere. Hence, we have for almost every $\theta$ that
\begin{equation}
 \dsdft{\Xb} =\left[\begin{array}{cc}
               \Pi\dsdft{(X_1/\sigma_1,\dots,X_{L'}/\sigma_{L'})}\Pi & 0\\0 & 0
              \end{array}\right]
\end{equation}
where $0$ denotes an all-zero matrix of appropriate size. We thus have $\rank{\dsdft{\Xb}}=\rank{\dsdft{(X_1/\sigma_1,\dots,X_{L'}/\sigma_{L'})}}$ for almost every $\theta$.
\end{IEEEproof}

\begin{lem}\label{lem:thm:approxbound}
	Let $\Xb$ be an $\ell$-variate, real-valued, Gaussian vector with mean vector $\bfmu_{\Xb}$ and covariance matrix $\covmat{\Xb}$. Let $\Wb\triangleq[\Xb]_m+\Ub$, where $\Ub$ is an $\ell$-variate vector, independent of $\Xb$, with components independently and uniformly distributed on $[0,1/m)$.
    Then,
	\begin{equation}
	\frac{\kld{\pdf{\Wb}}{\pdfg{\Wb}}}{\ell} \le \frac{1}{2}\log\left(2 \pi \left(1+\frac{1}{12}\right)\right) + \frac{75}{2} + \frac{24}{\pi}.
	\end{equation}
\end{lem}

\begin{IEEEproof}
 By \cite[Th.~23.6.14]{Lapidoth_DigitalCommunication}, $\Xb=\trans{(X_1,\ldots,X_\ell)}$ can be written as
\begin{equation}
\label{eq:XbNb}
\Xb = A \Xb' + \bfmu_{\Xb}
\end{equation}
where $\Xb'$ is an $\ell'$-dimensional, zero-mean, Gaussian vector ($\ell'\leq \ell$) with independent components whose variances are the nonzero eigenvalues of $\covmat{\Xb}$ and where $A$ is an $\ell\times \ell'$ matrix satisfying $\trans{A} A=I_{\ell'}$. We use the data processing inequality, the chain rule for relative entropy, and the fact that $\Xb'$ is Gaussian, to obtain
\begin{IEEEeqnarray}{lCl}
\IEEEeqnarraymulticol{3}{l}{ \kld{\pdf{\Wb}}{\pdfg{\Wb}}} \nonumber\\
\quad & \le & \kld{\pdf{\Wb,\Xb'}}{\pdfg{\Wb,\Xb'}}\nonumber\\
& \le & \kld{\pdf{\Xb'}}{\pdfg{\Xb'}} + \int \kld{\pdf{\Wb|\Xb'=\mathbf{x}'}}{\pdfg{\Wb|\Xb'=\mathbf{x}'}} \pdf{\Xb'}(\mathbf{x}') \diff \mathbf{x}' \nonumber\\
& = & \int \kld{\pdf{\Wb|\Xb'=\mathbf{x}'}}{\pdfg{\Wb|\Xb'=\mathbf{x}'}} \pdf{\Xb'}(\mathbf{x}') \diff \mathbf{x}'\label{eq:klddpi}
\end{IEEEeqnarray}
where $\pdfg{\Wb,\Xb'}$ denotes the PDF of a Gaussian vector with the same mean vector and covariance matrix as $(\Wb,\Xb')$, and
\begin{align}
\pdf{\Wb|\Xb'=\mathbf{x}'}(\mathbf{w}) &\triangleq \frac{\pdf{\Wb,\Xb'}(\mathbf{w},\mathbf{x}')}{\pdf{\Xb'}(\mathbf{x}')}\\
 \pdfg{\Wb|\Xb'=\mathbf{x}'}(\mathbf{w}) &\triangleq \frac{\pdfg{\Wb,\Xb'}(\mathbf{w},\mathbf{x}')}{\pdfg{\Xb'}(\mathbf{x}')}.
\end{align}

To evaluate the relative entropy on the RHS of \eqref{eq:klddpi}, we first note that, given $\Xb$, the random vector $\Wb$ is uniformly distributed on an $\ell$-dimensional cube of length $\frac{1}{m}$. Since $\Xb$ can be obtained from $\Xb'$ via \eqref{eq:XbNb}, the conditional PDF of $\Wb$ given $\Xb'=\mathbf{x}'$ is
\begin{equation}
\pdf{\Wb|\Xb'=\mathbf{x}'}(\mathbf{w}) =  m^\ell \mathbf{1}\{\qRV{\mathbf{w}}{m} = \qRV{A\mathbf{x}'+\bfmu_{\Xb}}{m}\}.
\end{equation}
Consequently, denoting $\mathbf{z}=\qRV{A\mathbf{x}'+\bfmu_{\Xb}}{m}$,
\begin{IEEEeqnarray}{lCl}
\kld{\pdf{\Wb|\Xb'=\mathbf{x}'}}{\pdfg{\Wb|\Xb'=\mathbf{x}'}} & = & \log\left( m^\ell \sqrt{(2\pi)^\ell\det\covmat{\Wb|\Xb'} }\right)\nonumber\\
\IEEEeqnarraymulticol{3}{r}{ {} + \frac{m^\ell}{2} \int_{\dom{C}(\mathbf{z},1/m)} \!\!\!\!\!\trans{(\mathbf{w}-\bfmu_{\Wb|\Xb'=\mathbf{x}'})}\covmat{\Wb|\Xb'}^{-1}(\mathbf{w}-\boldsymbol{\mu}_{\Wb|\Xb'=\mathbf{x}'})  \diff \mathbf{w}} \nonumber\\  \label{eq:kld_cond}
\end{IEEEeqnarray}
where $\bfmu_{\Wb|\Xb'=\mathbf{x}'}$ and $\covmat{\Wb|\Xb'}$ denote the conditional mean and the conditional covariance matrix of $\Wb$ given $\Xb'=\mathbf{x}'$. These can be computed as \cite[Th.~23.7.4]{Lapidoth_DigitalCommunication}
\begin{align}
 \bfmu_{\Wb|\Xb'=\mathbf{x}'} & = \expec{\Wb} + \covmat{\Wb \Xb'}\covmat{\Xb'}^{-1} \mathbf{x}' \label{eq:mu_WbNb} \\
 \covmat{\Wb|\Xb'} & = \covmat{\Wb} - \covmat{\Wb \Xb'}\covmat{\Xb'}^{-1}\trans{\covmat{\Wb \Xb'}} \label{eq:C_WbNb}
\end{align}
where $\covmat{\Wb \Xb'}$ denotes the cross-covariance matrix of $\Wb$ and $\Xb'$, and $\covmat{\Wb}$ and $\covmat{\Xb'}$ denote the covariance matrices of $\Wb$ and $\Xb'$, respectively.

Defining $\Zb\triangleq [\Xb]_m$, we have $\Wb=\Zb+\Ub$. Since $\Ub$ is independent of $\Xb$, the cross-covariance matrix of $\Wb$ and $\Xb$ is equal to the cross-covariance matrix of $\Zb$ and $\Xb$. Bussgang's theorem~\cite[eq.~(20)]{bussgang52} yields \mbox{$\covt{Z_jX_i} = a_{j}  \covt{X_jX_i}$}, where $a_j$ is defined in \eqref{eq:bounda}. Hence, if $\Lambda_\mathbf{a}$ is a diagonal matrix with $\mathbf{a}=(a_1,\dots,a_\ell)$ on the main diagonal, then $\covmat{\Zb\Xb}=\Lambda_\mathbf{a} \covmat{\Xb}$.  From~\eqref{eq:XbNb} we get $\covmat{\Xb}=A\covmat{\Xb'}\trans{A}$ and $\covmat{\Wb\Xb'}=\covmat{\Wb\Xb}A$, hence
\begin{equation}
 \covmat{\Wb\Xb'}=\covmat{\Wb\Xb}A=\covmat{\Zb\Xb}A=\Lambda_\mathbf{a} \covmat{\Xb}A=\Lambda_\mathbf{a}A\covmat{\Xb'}.
\end{equation}
Together with \eqref{eq:mu_WbNb} and \eqref{eq:C_WbNb}, this yields
\begin{IEEEeqnarray}{rCl}
 \bfmu_{\Wb|\Xb'=\mathbf{x}'} & = & \expec{\Wb} + \Lambda_\mathbf{a} A \mathbf{x}' \label{eq:mu_WbNb_2}\\
 \covmat{\Wb|\Xb'} & = & \covmat{\Wb} - \Lambda_\mathbf{a}\covmat{\Xb}\Lambda_\mathbf{a}. \label{eq:cov_wbnb_2}
\end{IEEEeqnarray}

Combining \eqref{eq:mu_WbNb_2}  with \eqref{eq:XbNb}, and using the triangle inequality, we upper-bound each component of $\mathbf{w}-\bfmu_{\Wb|\Xb'=\mathbf{x}'}$ as
 \begin{multline}
 \left|w_{j} - \expec{W_{j}} - a_j(x_{j}-\mu_j)\right|   \leq  |z_{j}-x_{j}| + \left|u_{j} - \expec{U_{j}}\right|\\
 + \left|\expec{Z_{j}}-\mu_j\right| + |1-a_j| |x_{j}-\mu_j|. \label{eq:difference_ISIT}
 \end{multline}
The first and the third term on the RHS of \eqref{eq:difference_ISIT} are both upper-bounded by $\frac{1}{m}$, and the second term is upper-bounded by $\frac{1}{2m}$. From \eqref{eq:boundBussgang} in Lemma~\ref{lem:thm:bussgang}, we get that the term $|1-a_j|$ is upper-bounded by $1/m\sqrt{2/\pi\sigma_j^2}$, where $\sigma_j^2$ is the variance of $X_j$. We thus obtain
\begin{equation}
\label{eq:bound_norm}
\|\mathbf{w}-\bfmu_{\Wb|\Xb'=\mathbf{x}'}\|_2^2 \leq \frac{1}{m^2}\left(\frac{25\ell}{2} + \frac{4}{\pi}\sum_{j=1}^\ell \frac{(x_j-\mu_j)^2}{\sigma_j^2} \right).
\end{equation}

We next note that, since $\Wb=\Zb+\Ub$, and since $\Ub$ is independent from $\Zb$ and i.i.d. on $[0,1/m)$,
\begin{equation}
\label{eq:covmat_lemma1}
\covmat{\Wb|\Xb'} = \covmat{\Zb} - \Lambda_\mathbf{a}\covmat{\Xb}\Lambda_\mathbf{a} + \frac{1}{12m^2} I_{\ell}.
\end{equation}
It can be shown that $\covmat{\Zb} - \Lambda_\mathbf{a} \covmat{\Xb}\Lambda_\mathbf{a}$ is the conditional covariance matrix of $\Zb$ given $\Xb'$, hence it is positive semidefinite.\footnote{Indeed, we have $\covmat{\Zb\Xb}=\covmat{\Wb\Xb}$ and, by \eqref{eq:XbNb}, $\covmat{\Zb\Xb'}=\covmat{\Zb\Xb}A$. Replacing in \eqref{eq:C_WbNb} $\Wb$ by $\Zb$, and repeating the steps leading to \eqref{eq:cov_wbnb_2}, we obtain the desired result.} It follows that the smallest eigenvalue of $\covmat{\Wb|\Xb'}$ is lower-bounded by $\frac{1}{12 m^2}$. Together with \eqref{eq:bound_norm}, this yields for the second term on the RHS of \eqref{eq:kld_cond}
\begin{IEEEeqnarray}{lCl}
\IEEEeqnarraymulticol{3}{l}{\frac{m^\ell}{2} \int_{\dom{C}(\mathbf{z},1/m)} \trans{(\mathbf{w}-\bfmu_{\Wb|\Xb'=\mathbf{x}'})}\covmat{\Wb|\Xb'}^{-1}(\mathbf{w}-\bfmu_{\Wb|\Xb'=\mathbf{x}'})  \diff \mathbf{w}} \nonumber\\
\quad & \leq &  6 m^{\ell+2} \frac{1}{m^\ell} \frac{1}{m^2}\left(\frac{25\ell}{2} + \frac{4}{\pi}\sum_{j=1}^\ell \frac{(x_j-\mu_j)^2}{\sigma_j^2}\right)\nonumber\\
 & = & \frac{75\ell}{2} + \frac{24}{\pi}\sum_{j=1}^\ell \frac{(x_j-\mu_j)^2}{\sigma_j^2}. \label{eq:split_1}
\end{IEEEeqnarray}

To upper-bound the first term on the RHS of \eqref{eq:kld_cond}, we use that \eqref{eq:covmat_lemma1} combined with Lemma~\ref{lem:thm:bussgang} implies that every diagonal element of $\covmat{\Wb|\Xb'}$ is given by
 \begin{multline}
 \expec{(Z_j-\expec{Z_j})^2} - a_j^2\sigma_j^2 + \frac{1}{12m^2}
  =  -(1-a_j)^2 \sigma_j^2 \\ {} + \expec{(X_j-\mu_j-Z_j+\expec{Z_j})^2} + \frac{1}{12m^2}. \label{eq:dimrateproof:moreslowly}
 \end{multline}
The first term on the RHS of \eqref{eq:dimrateproof:moreslowly} is negative, and the second term is upper-bounded by $ \expec{(X_j-Z_j)^2}\le 1/m^2$. Hence, every element on the main diagonal of $\covmat{\Wb|\Xb'}$ is upper-bounded by $\frac{1+1/12}{m^2}$. It thus follows from Hadamard's inequality that
\begin{equation}
\label{eq:split_2}
\log\left( m^\ell \sqrt{(2\pi)^\ell\det\covmat{\Wb|\Xb'} }\right) \leq \frac{\ell}{2} \log\left(2 \pi \left(1+\frac{1}{12}\right)\right).
\end{equation}
Combining \eqref{eq:split_1} and \eqref{eq:split_2} with \eqref{eq:kld_cond} and \eqref{eq:klddpi} yields
\begin{IEEEeqnarray}{lCl}
\IEEEeqnarraymulticol{3}{l}{\kld{\pdf{\Wb_1^\ell}}{\pdfg{\Wb_1^\ell}}} \nonumber\\
& \leq & \frac{\ell}{2} \log\left(2 \pi \left(1+\frac{1}{12}\right)\right) + \frac{75\ell}{2} + \frac{24}{\pi}\sum_{j=1}^\ell \frac{\expec{(X_j-\mu_j)^2}}{\sigma_j^2}\nonumber\\
& = & \ell\left(\frac{1}{2}\log\left(2 \pi \left(1+\frac{1}{12}\right)\right) + \frac{75}{2} + \frac{24}{\pi} \right)
\end{IEEEeqnarray}
and completes the proof.
\end{IEEEproof}

\section{Spectral Distribution Function of $\{\qRVm{\Xb_t}\}$}
\label{app:bussgang}
Let $\{\Xb_t\}$ be a stationary, $L$-variate, Gaussian process with mean vector $\bfmu=\trans{(\mu_1,\ldots,\mu_L)}$ and SDF $\sdf{\Xb}$. Let $\{\Zb_t\}$ and $\{\Nb_t\}$ be defined as $Z_{i,t}\triangleq \qRV{X_{i,t}}{m}$ and $N_{i,t}\triangleq X_{i,t}-\qRVm{X_{i,t}}$, respectively. For every pair $i,j=1,\dots,L$, we have
\begin{align}
 \covt{N_iN_j} = & \covt{X_iX_j} + \covt{Z_iZ_j}
 \nonumber\\ &{} 
 - \covt{X_iZ_j} - \covt{Z_iX_j}.
\end{align}
Bussgang's theorem~\cite[eq.~(20)]{bussgang52} further yields that $\covt{X_iZ_j} = \cov{Z_jX_i}(-\tau)= a_{j}  \covt{X_iX_j}$, where $a_j$ is defined in \eqref{eq:bounda}. Consequently,
 \begin{align}
  \covt{N_iN_j} 
  &=\covt{X_iX_j} +  \covt{Z_iZ_j}
 \notag\\
  &\quad 
 - a_j\covt{X_iX_j} - a_i\cov{X_jX_i}(-\tau)\notag\\
  &=  (1-a_j-a_i)\covt{X_iX_j} + \covt{Z_iZ_j}.
 \end{align}
 Since the SDF is fully determined by the covariance structure of a process \cite[Th.~1, p.~206]{gihmanskorohod80}, we obtain~\eqref{eq:SDMSum}.  
 
To prove~\eqref{eq:boundNoise}, namely,
\begin{equation}
\int_{-1/2}^{1/2} \diff\sdft{N_i} \leq \frac{1}{m^2}
\end{equation}
we note that
\begin{equation}
\int_{-1/2}^{1/2} \diff\sdft{N_i} = \expec{(X_{i,t}-Z_{i,t})^2} - (\mu_i-\expec{Z_{i,t}})^2.
\end{equation}
Since $|X_{i,t}-Z_{i,t}|\leq\frac{1}{m}$ and $(\mu_i-\expec{Z_{i,t}})^2\geq 0$, the claim~\eqref{eq:boundNoise} follows.

It remains to prove \eqref{eq:boundBussgang}, namely,
 \begin{equation}
|1-a_i| \le \frac{1}{m} \sqrt{\frac{2}{\pi \sigma_i^2}}.
 \end{equation}
Set $f(\alpha)\triangleq \frac{\alpha}{\sigma_i}\e{-\alpha^2/2\sigma_i^2}$, $\alpha\in\mathbb{R}$. We have
 \begin{IEEEeqnarray}{lCl}
  a_i & = & \frac{\expec{(X_{i,t}-\mu_i)(Z_{i,t}-\expec{Z_{i,t}})}}{\sigma_i^2} \nonumber\\
  & = & \frac{\expec{(X_{i,t}-\mu_i)Z_{i,t}}}{\sigma_i^2} \nonumber\\
  & = & \frac{1}{\sqrt{2\pi}\sigma_i^2} \sum_i \frac{i}{m} \int_{\frac{i}{m}}^{\frac{i+1}{m}} f(\alpha)\diff\alpha.
 \end{IEEEeqnarray}
 Furthermore,
 \begin{equation}
 \frac{1}{\sqrt{2\pi}\sigma_i^2}  \int_{-\infty}^\infty \alpha f(\alpha)\diff\alpha  = 1.
 \end{equation}
 It follows that
 \begin{equation}
 1 - a_i = \frac{1}{\sqrt{2\pi}\sigma_i^2} \sum_i \int_{\frac{i}{m}}^{\frac{i+1}{m}} \left(\alpha-\frac{i}{m}\right) f(\alpha)\diff\alpha.
 \end{equation}
Since $|\alpha-i/m|\le 1/m$ for $\alpha\in[i/m,(i+1)/m]$, this yields
 \begin{align}
  |1-a_i| &\le\frac{1}{\sqrt{2\pi}\sigma_i^2} \sum_{i} \int_{\frac{i}{m}}^{\frac{i+1}{m}} \left|\alpha-\frac{i}{m}\right| |f(\alpha)|\diff\alpha\notag\\
  & \le \frac{1}{m} \frac{1}{\sqrt{2\pi}\sigma_i^2} \int_{-\infty}^{\infty} |f(\alpha)|\diff\alpha\notag\\
  & = \frac{1}{m} \sqrt{\frac{2}{\pi \sigma_i^2}}.
 \end{align}
This proves \eqref{eq:boundBussgang} and concludes the proof of Lemma~\ref{lem:thm:bussgang}.

\section{Proof of Theorem~\ref{thm:rate:proper}}\label{proofs:rate:proper}
Let $\{\Zb_t\}$ be a stationary, $L$-variate, complex-valued process with matrix-valued SDF $\sdf{\Zb}$. Let the \emph{real composite process} $\{\hat{\Xb}_t\}$ be defined as $\hat{\Xb}_t\triangleq \trans{[\mathfrak{Re}(\trans{\Zb_t}),\mathfrak{Im}(\trans{\Zb_t})]}$. That is, $\hat{\Xb}_t$ is obtained by stacking the real part of $\Zb_t$ on top of the imaginary part of $\Zb_t$. Further let the \emph{augmented process} $\{\hat{\Zb}_t\}$ be defined as $\hat{\Zb}_t\triangleq \trans{[\trans{\Zb_t}, \herm{\Zb}_t]}$. Clearly, $\hat{\Xb}_t$ and $\hat{\Zb}_t$ satisfy $\hat{\Zb}_t=T\hat{\Xb}_t$, where
\begin{equation}
T \triangleq \left[\begin{array}{cc} I_L & \imath I_L \\ I_L & -\imath I_L \end{array}\right]
\end{equation}
is unitary up to a factor of $2$, i.e., $T \herm{T}=\herm{T} T = 2 I_L$. The matrix-valued autocovariance function of $\{\hat{\Zb}_t\}$ reads
\begin{equation}\label{eq:rate:augmentedMatrix:cov}
 \covt{\hat{\Zb}} = \left[\begin{array}{cc}
                                  \covt{\Zb} & \pcovt{\Zb}\\
                                  \conj{\pcov{\Zb}}(\tau) & \conj{\cov{\Zb}}(\tau)
                                 \end{array}
 \right]
\end{equation}
where $\pcov{\Zb}$ denotes the pseudo-autocovariance function of $\{\Zb_t\}$. The corresponding matrix-valued SDF is given by
\begin{equation}\label{eq:rate:augmentedMatrix:sdf}
  \sdft{\hat{\Zb}} = \left[\begin{array}{cc}
                                  \sdft{\Zb} & \psdf{\Zb}(\theta)\\
                                  -\conj{\psdf{\Zb}} (-\theta)& -\conj{\sdf{\Zb}}(-\theta)
                                 \end{array}
 \right]
\end{equation}
where $\psdf{\Zb}$ satisfies
\begin{equation}
\pcov{\Zb}(\tau) = \int_{-1/2}^{1/2} e^{-\imath 2 \pi \tau \theta}\diff\psdf{\Zb}(\theta), \quad \tau\in\mathbb{Z}.
\end{equation}
The autocovariance functions and SDFs of $\{\hat{\Xb}_t\}$ and $\{\hat{\Zb}_t\}$ are related via
\begin{align}
\covt{\hat{\Zb}} & = T\covt{\hat{\Xb}}\herm{T} \\
\sdft{\hat{\Zb}} & =T\sdft{\hat{\Xb}}\herm{T}. \label{eq:complex_sdf}
\end{align}

By definition, $\infodimlu{\{\Zb_t\}}=\infodimlu{\{\hat{\Xb}_t\}}$. It thus follows from Theorem~\ref{thm:rate:Gauss} that
\begin{equation}
\label{eq:Bla_Th10}
 \infodimlu{\{{\Zb}_t\}} = \infodimlu{\{\hat{\Xb}_t\}} \leq \int_{-1/2}^{1/2}\rank{\dsdft{\hat{\Xb}}}\diff\theta.
\end{equation}
Since left or right multiplication by a nonsingular matrix leaves the rank unchanged, we obtain from \eqref{eq:complex_sdf} that the rank of $\dsdft{\hat{\Xb}}$ is equal to the rank of $\dsdft{\hat{\Zb}}$. Furthermore, by \eqref{eq:rate:augmentedMatrix:sdf}, the rank of $\dsdft{\hat{\Zb}}$ is upper-bounded by the rank of $\dsdft{\Zb}$ plus the rank of $\conj{({\dsdf{\Zb}})}(-\theta)$ \cite[Th.~1]{Lundquist_Rank}. Consequently,
\begin{align}
 \infodimlu{\{{\Zb}_t\}} & \leq  \int_{-1/2}^{1/2}\rank{\dsdft{\Zb}}+\rank{\conj{({\dsdf{\Zb}})}(-\theta)}\diff\theta\nonumber\\
  &= 2\int_{-1/2}^{1/2}\rank{\dsdft{\Zb}} \diff\theta \label{eq:Bla_Th10_2}
\end{align}
where the second step follows because complex conjugation does not affect the rank.

If $\{{\Zb}_t\}$ is Gaussian, then \eqref{eq:Bla_Th10} holds with equality by Theorem~\ref{thm:rate:Gauss}. If $\{{\Zb}_t\}$ is, in addition, proper then $\pcov{\Zb}(\tau)=0$, so the derivative of $\psdf{\Zb}$ is zero almost everywhere. Hence, the derivative of $\sdf{\hat{\Zb}}$ becomes block diagonal almost everywhere and its rank equals the sum of the ranks of its diagonal elements. We conclude that, if $\{\Zb_t\}$ is proper Gaussian, then \eqref{eq:Bla_Th10_2} holds with equality. This proves Theorem~\ref{thm:rate:proper}.

\section{Appendix to Section~\ref{sec:poorrate}}
\subsection{Proof of Theorem~\ref{thm:dimrates}}
\label{app:thm:dimrates}
For every $m=2,3,\ldots$ and $k=1,2,\ldots$ we have
 \begin{align}
  \ent{\qRV{X_1}{m}|X_{-\infty}^0} & \le \ent{\qRV{X_k}{m}|\qRV{X^{k-1}_{-\infty}}{m}} \nonumber\\
  & \le \ent{\qRV{X_k}{m}|\qRV{X_1^{k-1}}{m}}\label{eq:dimrateproof:block}
 \end{align}
by stationarity; and because conditioning reduces entropy and, conditioned on $X_{-\infty}^0$, $\qRV{X_1}{m}$ is independent of $\qRV{X_{-\infty}^0}{m}$. Note that, by \eqref{eq:entrate_alt} and stationarity,
\begin{equation}
\label{eq:dimrateproof:ent}
\ent{\qRV{X_k}{m}|\qRV{X_{-\infty}^{k-1}}{m}}=\entrate{\{\qRV{X_t}{m}\}}.
\end{equation}
Thus, dividing \eqref{eq:dimrateproof:block} by $\log m$ and taking first the limit over $m$ and then the limit over $k$ yields
\begin{equation}
\label{eq:dimrateproof:inequality}
\infodimlu{X_1|X_{-\infty}^0} \le \infodimlu{\{X_t\}} \leq \dimratelu.
\end{equation}
This proves \eqref{eq:dimrates}.

We next bound the difference $\dimratelu-\infodimlu{\{X_t\}}$. By \eqref{eq:dimrateproof:ent}, we have
 \begin{multline}
 \label{eq:dimrateproof:missing}
 \ent{\qRV{X_k}{m}|\qRV{X_1^{k-1}}{m}} - \entrate{\{\qRV{X_t}{m}\}}\\
 = I(\qRV{X_k}{m};\qRV{X_{-\infty}^0}{m}|\qRV{X_1^{k-1}}{m}).
 \end{multline}
Dividing \eqref{eq:dimrateproof:missing} by $\log m$ and taking first the limit over $m$ and then the limit over $k$ yields
  \begin{IEEEeqnarray}{lCl}
\IEEEeqnarraymulticol{3}{l}{\limk \varliminf_{m\to\infty} \frac{I(\qRV{X_k}{m};\qRV{X_{-\infty}^0}{m}|\qRV{X_1^{k-1}}{m})}{\log m}} \nonumber\\
\quad & \leq & \dimratelu - \infodimlu{\{X_t\}} \nonumber\\
 & \leq & \limk \varlimsup_{m\to\infty} \frac{I(\qRV{X_k}{m};\qRV{X_{-\infty}^0}{m}|\qRV{X_1^{k-1}}{m})}{\log m}.
  \end{IEEEeqnarray}
This concludes the proof of Theorem~\ref{thm:dimrates}.

\subsection{Proof of Corollary~\ref{cor:dimrates}}
\label{proof:cor_dimrates}
Suppose there exists a nonnegative $n$ such that
\begin{equation}\label{eq:pinCondProof}
 \mutinf{X_1^k;X_{-\infty}^{-n}} < \infty, \quad k=1,2,\ldots
\end{equation}
We first show that
\begin{equation}
\label{eq:cor_dimrates_1}
 \mutinf{\qRV{X_k}{m};\qRV{X_{-\infty}^0}{m}|\qRV{X_1^{k-1}}{m}}
\leq \frac{\mutinf{\qRV{X_1^k}{m};\qRV{X_{-\infty}^0}{m}}}{k}.
\end{equation}
In a second step, we then show that \eqref{eq:pinCondProof} implies that
\begin{equation}
\label{eq:cor_dimrates_1.5}
\lim_{k\to\infty} \varlimsup_{m\to\infty} \frac{I(\qRV{X_1^k}{m};\qRV{X_{-\infty}^0}{m})}{k\log m} = 0
\end{equation}
which together with \eqref{eq:cor_dimrates_1} and \eqref{eq:thm_dimrates} demonstrates that $\infodimlu{\{X_t\}}=\dimratelu$, thus proving Corollary~\ref{cor:dimrates}.

To prove \eqref{eq:cor_dimrates_1}, we use the chain rule, stationarity, and the fact that conditioning reduces entropy, to obtain
\begin{IEEEeqnarray}{lCl}
\IEEEeqnarraymulticol{3}{l}{I(\qRV{X_1^k}{m};\qRV{X_{-\infty}^0}{m})} \nonumber\\
\quad & = & \sum_{\ell=1}^k \Bigl[\ent{\qRV{X_{\ell}}{m}|\qRV{X_1^{\ell-1}}{m}}-\ent{\qRV{X_{\ell}}{m}|\qRV{X_{-\infty}^{\ell-1}}{m}}\Bigr] \nonumber\\
& \geq & \sum_{\ell=1}^k \Bigl[\ent{\qRV{X_{k}}{m}|\qRV{X_1^{k-1}}{m}}-\ent{\qRV{X_{k}}{m}|\qRV{X_{-\infty}^{k-1}}{m}}\Bigr] \nonumber\\
& = & k I(\qRV{X_k}{m};\qRV{X_{-\infty}^0}{m}|\qRV{X_1^{k-1}}{m}). \label{eq:cor_dimrates_2}
\end{IEEEeqnarray}
Having obtained \eqref{eq:cor_dimrates_1}, we next show that \eqref{eq:pinCondProof} implies \eqref{eq:cor_dimrates_1.5}. Indeed,
\begin{IEEEeqnarray}{lCl}
\IEEEeqnarraymulticol{3}{l}{\varlimsup_{k\to\infty} \varlimsup_{m\to\infty} \frac{I(\qRV{X_1^k}{m};\qRV{X_{-\infty}^0}{m})}{k\log m}} \nonumber\\
\quad &\le&  \varlimsup_{k\to\infty} \limmsup \frac{\mutinf{\qRV{X_1^k}{m};\qRV{X_{-\infty}^{-n}}{m}}}{k\log m}
\nonumber \\
 && {}
+ \varlimsup_{k\to\infty} \limmsup \frac{\mutinf{\qRV{X_1^k}{m};\qRV{X_{-n+1}^0}{m}|\qRV{X_{-\infty}^{-n}}{m}}}{k\log m}\nonumber\\
 &\le & \varlimsup_{k\to\infty} \limmsup \frac{\mutinf{X_1^k;X_{-\infty}^{-n}}}{k\log m} \nonumber\\
 & &{} +\varlimsup_{k\to\infty} \limmsup \frac{\ent{\qRV{X_{-n+1}^0}{m}}}{k\log m} \label{eq:dimrateproof:last}
\end{IEEEeqnarray}
where $n$ is a nonnegative integer satisfying \eqref{eq:pinCondProof}. Here, the first inequality follows from the chain rule; the second inequality follows from the data processing inequality and by upper-bounding the second mutual information by $ \ent{\qRV{X_{-n+1}^0}{m}}$.

The first limit on the RHS of \eqref{eq:dimrateproof:last} is zero because, by assumption, $\mutinf{X_1^k;X_{-\infty}^{-n}}<\infty$. The second limit on the RHS of \eqref{eq:dimrateproof:last} can be written as $\varlimsup_{k\to\infty} \infodimu{X_{-n+1}^0}/k$, which is zero because, by Lemma~\ref{lem:dim:finite}, $\infodimu{X_{-n+1}^0}$ is bounded in $k$. This proves \eqref{eq:cor_dimrates_1.5} and concludes the proof of Corollary~\ref{cor:dimrates}.

\subsection{Proof of Lemma~\ref{lem:TobiPredictionError}}\label{proof:TobiPredictionError}
 Since $\{X_t\}$ is Gaussian, the conditional mean of $X_k$ given $X_0,\dots,X_{k-1}$ can be written as
 \begin{equation}
  \expec{X_k|X_0,\dots,X_{k-1}} = \sum_{\ell=1}^k \alpha_\ell X_{k-\ell}
 \end{equation}
 for some coefficients $\alpha_1,\ldots,\alpha_k$.\footnote{More precisely, the coefficients correspond to the LMMSE estimator for estimating $X_k$ from $X_0,\ldots,X_{k-1}$. The LMMSE estimator always exists, even though it is not necessarily unique.}
The conditional variance $\sigma_k^2$ is thus given by (see, e.g., \cite[Sec.~10.6]{grenanderszego58})
 \begin{multline}
 \label{eq:szegoe}
  \sigma_k^2 = \expec{\left(X_k-\sum_{\ell=1}^k \alpha_\ell X_{k-\ell}\right)^2}
 \\ = \int_{-\frac{1}{2}}^{\frac{1}{2}} \left|1-\sum_{\ell=1}^k \alpha_\ell \e{-\imath 2\pi \ell \theta}\right|^2 \diff \sdft{X}.
 \end{multline}
The function
\begin{equation}
g(\theta) = 1-\sum_{\ell=1}^k \alpha_\ell \e{-\imath 2\pi \ell \theta},\quad -1/2\leq\theta\leq 1/2
\end{equation}
is analytic on the closed interval $[-1/2,1/2]$, hence it is either constant or it has at most finitely many zeros in $[-1/2,1/2]$. Moreover, $g$ cannot be the all-zero function, as can be argued by contradiction. Indeed, suppose there exist $\alpha_1,\ldots,\alpha_k$ such that $g(\theta)=0$ for all $\theta$. Then, by \eqref{eq:szegoe}, we have $\sigma_k^2=0$ irrespective of $\sdf{X}$. In other words, we can find a linear estimator that perfectly predicts $X_k$ from $X_0,\ldots,X_{k-1}$ irrespective of the SDF of $\{X_t\}$. This is clearly a contradiction, since even the best predictor yields $\sigma_k^2=\sigma^2$ for an i.i.d., zero-mean, variance-$\sigma^2$, Gaussian process, i.e., when $\sdfdt{X}=\sigma^2$. Thus, the set $\dom{Z}\triangleq\{\theta : g(\theta)=0\}$ is finite and has therefore Lebesgue measure zero.

Since $|g(\theta)|^2=0$ for $\theta\in\dom{Z}$, we have
\begin{equation}
 \sigma_k^2=\int_{\dom{Z}^\mathsf{c}} |g(\theta)|^2 \diff \sdft{X}.
\end{equation}
Since furthermore $|g(\theta)|^2>0$ for $\theta\in\dom{Z}^\mathsf{c}$, we have $\sigma_k^2=0$ only if
\begin{equation}
 \int_{\dom{Z}^\mathsf{c}} \diff \sdft{X} = 0.
\end{equation}
This implies that $\dsdft{X}=0$ for all $\theta\in\dom{Z}^\mathsf{c}$. Hence, the set of harmonics $\theta$ for which $\dsdft{X}>0$ is contained in $\dom{Z}$. The proof is completed by the monotonicity of measures and the fact that $\dom{Z}$ has Lebesgue measure zero.

\section*{Acknowledgment}
Fruitful discussions with Amos Lapidoth are gratefully acknowledged. The authors further wish to thank the Associate Editor Matthieu Bloch and the anonymous referees for their valuable comments.


\begin{IEEEbiographynophoto}{Bernhard C. Geiger} (S'07--M'14--SM'19) was born in Graz, Austria, in 1984. He received the Dipl.-Ing. degree in electrical engineering (with distinction) and the Dr. techn. degree in electrical and information engineering (with distinction) from Graz University of Technology, Austria, in 2009 and 2014, respectively.

In 2009 he joined the Signal Processing and Speech Communication Laboratory, Graz University of Technology, as a Project Assistant and took a position as a Research and Teaching Associate at the same lab in 2010. He was a Senior Scientist and Erwin Schr\"odinger Fellow at the Institute for Communications Engineering, Technical University of Munich, Germany from 2014 to 2017 and a postdoctoral researcher at the Signal Processing and Speech Communication Laboratory, Graz University of Technology, Austria from 2017 to 2018. He is currently Senior Researcher at Know-Center GmbH, Graz, Austria. His research interests cover information theory for machine learning and information-theoretic model reduction for Markov chains and hidden Markov models.
\end{IEEEbiographynophoto}

\vfill

\newpage

\begin{IEEEbiographynophoto}{Tobias Koch} (S'02--M'09--SM'16) is a Visiting Professor and Ram\'on y Cajal Research Fellow with the Signal Theory and Communications Department of Universidad Carlos III de Madrid (UC3M). He received the M.Sc.\ degree in electrical engineering (with distinction) in 2004 and the Ph.D.\ degree in electrical engineering in 2009, both from ETH Zurich, Switzerland. From June 2010 until May 2012 he was a Marie Curie Intra-European Research Fellow with the University of Cambridge, UK. He was also a research intern at Bell Labs, Murray Hill, NJ, USA in 2004, and the Universitat Pompeu Fabra (UPF), Barcelona, Spain, in 2007. He joined the Signal Processing Group of UC3M in June 2012. His research interests are in digital communication theory and information theory.

Dr.\ Koch received a Starting Grant from the European Research Council (ERC), a Ram\'on y Cajal Research Fellowship, a Marie Curie Intra-European Fellowship, a Marie Curie Career Integration Grant, and a Fellowship for Prospective Researchers from the Swiss National Science Foundation. In 2013--2016 he served as Vice Chair of the Spain Chapter of the IEEE Information Theory Society.
\end{IEEEbiographynophoto}

\end{document}